\documentclass{article}

\usepackage[english]{babel}

\usepackage[a4paper,top=2cm,bottom=2cm,left=2cm,right=2cm,marginparwidth=2cm]{geometry}
\usepackage{comment}
\usepackage{amsmath}
\usepackage{amssymb}
\usepackage{cite}
\usepackage{graphicx}
\usepackage[colorlinks=true, allcolors=blue]{hyperref}
\usepackage{authblk}
\usepackage{url}
\usepackage[figuresright]{rotating}
\usepackage{float}
\usepackage{multirow}
\usepackage{ulem}
\usepackage{todonotes}
\usepackage[title]{appendix}
\usepackage{longtable}
\usepackage[utf8]{inputenc}
\usepackage{caption}
\usepackage{epstopdf}
\usepackage{tikz}
\usetikzlibrary{shapes, arrows, positioning, fit, backgrounds, calc}
\usepackage{xcolor}  
\usepackage{listings}
\usepackage{array}
\usepackage{orcidlink}
\definecolor{sequenceRed}{RGB}{255,0,0}
\definecolor{sequenceBlue}{RGB}{0,0,255}
\definecolor{sequenceYellow}{RGB}{255,165,0}

\title{\textbf{Primer C-VAE: An interpretable deep learning primer design method to detect emerging virus variants}}
\author[1]{Hanyu Wang \orcidlink{0000-0002-9152-3899}}
\author[2]{Emmanuel K. Tsinda \orcidlink{0000-0003-1186-7657} \footnote{Currently affiliated with Center for Biomedical Innovation, MIT, Cambridge, MA 02139-4307, United States}}
\author[4]{Anthony J. Dunn}
\author[3]{Francis Chikweto \orcidlink{0000-0001-9404-0270}}
\author[4]{Alain B. Zemkoho \orcidlink{0000-0003-1265-4178}}
\affil[1]{School of Mathematics and Statistics, University of St Andrews, United Kingdom}
\affil[2]{Department of Virology, Tohoku University Graduate School of Medicine, Japan}
\affil[3]{Department of Biomedical Engineering, Tohoku University, Japan}
\affil[4]{School of Mathematical Sciences, University of Southampton, United Kingdom}

\begin{document}
\maketitle


\begin{abstract}

  \subsection*{Motivation:} Compared with next-generation sequencing (NGS), polymerase chain reaction (PCR) provides a more cost-effective and faster approach for detecting target organisms in both laboratory and field settings, where primer design is a critical step. In epidemiological surveillance of rapidly mutating viruses, designing effective primers is increasingly challenging. Traditional primer design workflows often require substantial manual intervention and may struggle to produce primers that remain specific across multiple strains within the same viral species (e.g., the Alpha and Delta variants of SARS-CoV-2). Similarly, for organisms with large and highly similar genomes (e.g., \textit{Escherichia coli} and \textit{Shigella flexneri}), designing primers that reliably discriminate between species is important but non-trivial. Therefore, more efficient and scalable primer design methods are needed.
  
  \subsection*{Results:} We introduce Primer C-VAE, a convolutional variational auto-encoder (C-VAE) that integrates a variational auto-encoder (VAE) framework with convolutional neural networks (CNNs), as the core model for identifying sequence variants. We then exploit features learned in the convolutional layers for downstream post-processing to derive candidate regions and generate variant-specific primers. Using SARS-CoV-2 as a case study, Primer C-VAE classified five variants (Alpha, Beta, Gamma, Delta, and Omicron) with 98\% accuracy on both the training and test sets, and generated primers for each variant. For most variants, the resulting primers appeared in more than 95\% of target sequences and less than 5\% of non-target sequences; Omicron showed moderately lower specificity (~80\% and ~20\%) owing to its greater genetic diversity. These primers showed good performance in \textit{in silico} PCR tests. For the Alpha, Delta, and Omicron variants, the primer pairs produced amplicons shorter than 200~bp, enabling their use in downstream qPCR assay development. In addition, Primer C-VAE successfully generated effective primers for longer genomes, including \textit{E. coli} and \textit{S. flexneri}.

  \subsection*{Conclusion:} Primer C-VAE is an interpretable deep-learning-based primer design approach for developing highly specific primer pairs for target organisms. It provides a flexible, semi-automated, and reliable workflow that is applicable across a range of sequence lengths and degrees of genome completeness. The method also supports downstream quantification applications (e.g., qPCR) and can be applied to a broad range of organisms, including those with large and highly similar genomes.
\end{abstract}


\section{Introduction} \label{section_introduction}

Primer design plays a crucial role in modern molecular biology. Through polymerase chain reaction (PCR), it enables a cost-effective and rapid approach for detecting organisms in genetic testing and research. As Bustin and Huggett emphasize, "primers are arguably the single most critical components of any PCR assay" \cite{Bustin_Huggett_2017}. Because primers determine where amplification begins and ends, their design directly influences assay specificity, sensitivity, and overall efficiency. However, for rapidly mutating viruses such as Severe Acute Respiratory Syndrome Coronavirus 2 (SARS-CoV-2), primer design becomes increasingly challenging as new variants continue to emerge. High mutation rates can necessitate the rapid development and updating of primer pairs to detect newly circulating variants, which in turn affects surveillance efforts and can inform clinical and public-health decision-making. Similarly, for closely related bacteria with high genomic similarity, such as \textit{Escherichia coli} (\textit{E. coli}) and \textit{Shigella flexneri} (\textit{S. flexneri}), designing primers that reliably discriminate between species is essential for accurate detection. Therefore, efficient and reliable primer design methods are critical for molecular diagnostics and downstream applications.

Currently, next-generation sequencing (NGS), also referred to as high-throughput sequencing, is widely used to identify emerging viral variants, including SARS-CoV-2 variants \cite{WHO_Headquarters}. However, implementing NGS typically requires sophisticated equipment and specialized expertise for both data generation and downstream analysis. Sanger sequencing offers a lower-cost alternative, but it remains time-consuming for genomes of approximately 30~kb, such as SARS-CoV-2 \cite{NGS_vs_Sanger}. In practice, this approach requires fragmenting the genome into multiple overlapping segments (approximately 700--900~bp each), sequencing each fragment individually, and then reassembling the full sequence using bioinformatics methods. For targeted detection when the organism (or gene) of interest is known, PCR assays provide a more economical and flexible alternative to sequencing-based workflows, which contributes to their widespread use in both research and clinical settings \cite{NGS_vs_qPCR, Johnson_2021}. Conventional PCR assays require a target sequence and a primer pair (forward and reverse) within a standard PCR reagent mixture \cite{Lorenz_2012}. The successful performance of these assays depends critically on primers that bind specifically and efficiently to the intended target region.

\subsection{Limitations of existing primer design methods}

Reliable PCR testing depends critically on primer design, which typically involves a forward and a reverse primer. A standard workflow begins by collecting homologous sequences that cover the target region. Multiple sequence alignment is then used to facilitate visual inspection and to identify conserved regions (typically 18--25 nucleotides) suitable for primer binding. Candidate primers are subsequently evaluated for key thermodynamic properties, including guanine-cytosine (GC) content, melting temperature, and potential secondary structures, to ensure efficient amplification and to minimise primer--dimer formation. The forward primer is designed to anneal to the antisense strand, whereas the reverse primer is designed downstream and anneals to the sense strand of the target DNA (or cDNA derived from RNA). For further details of this conventional methodology, see \cite{bustin2020ParametersSuccessfulPCR}.

In practice, however, primer design is considerably more complex than such schematic descriptions suggest. Rather than being a purely rule-based selection task, it often depends heavily on the designer's expertise in sequence alignment, specificity assessment, and artefact prevention (e.g., dimers and hairpins). For example, designing primers to detect the SARS-CoV-2 Alpha variant requires identifying an 18--25 nucleotide region within a \textasciitilde30,000-nucleotide genome that is both unique to the Alpha lineage and compatible with strict thermodynamic constraints \cite{lexa2001VirtualPCR}. Achieving adequate specificity typically requires extensive comparison against other SARS-CoV-2 variants and related coronaviruses, which can be time-consuming and cognitively demanding. These requirements, together with the need for sustained attention during manual screening, make the process prone to human error and inconsistency.

To reduce manual effort, tools such as Primer3 and Primer3Plus can rapidly generate candidate primers that satisfy basic design constraints, thereby decreasing repetitive work and some sources of human error. However, primers that meet \textit{in silico} criteria do not necessarily perform well in laboratory PCR experiments. Moreover, the specificity of automatically generated primers to a particular variant cannot be assumed and typically requires additional verification (e.g., database searches and experimental testing).

Another important limitation of commonly used tools is their restriction on input sequence length. For example, Primer3 processes sequences up to 10,000 base pairs, which makes it unsuitable for viruses such as SARS-CoV-2 with genomes exceeding 30,000~bp, and impractical for bacteria such as \textit{E. coli} with genomes on the order of millions of base pairs. As a result, existing pipelines often cannot be applied directly to long genomic sequences without substantial preprocessing or manual intervention.

Taken together, the limitations of existing primer design methods can be summarised as follows:

\begin{enumerate}
    \item Heavy reliance on specialist expertise and manual screening, resulting in time-intensive workflows that are vulnerable to human error.
    \item Limited ability to handle long genomic sequences, restricting applicability across diverse organisms.
    \item No inherent guarantee of primer specificity or experimental performance, necessitating additional \textit{in silico} and laboratory validation.
\end{enumerate}

To address these challenges, we propose a deep-learning-based, semi-automated primer design approach. Our method is designed to reduce manual screening, scale to long sequences, and generate more discriminative primers for closely related variants. This yields a streamlined and more robust workflow, which is particularly valuable for the detection of emerging pathogens and rapidly evolving viral lineages.

\subsection{Proposed method}

In this paper, we propose an interpretable deep-learning approach for primer design that generates both forward and reverse primers using a convolutional variational auto-encoder (C-VAE), which combines a variational auto-encoder (VAE) framework with convolutional neural networks (CNNs). We refer to this method as Primer C-VAE (Convolutional Variational Auto-Encoder for primer design), and show that it addresses several practical limitations of existing primer design workflows.

Primer C-VAE is designed for multiple primer design scenarios, including: (1) designing primers that discriminate between closely related variants within the same viral species (e.g., SARS-CoV-2 Alpha versus Delta), and (2) designing primers that distinguish between organisms with large and highly similar genomes (e.g., \textit{Escherichia coli} and \textit{Shigella flexneri}). In both settings, the goal is to produce primers that are not only thermodynamically feasible but also highly discriminative with respect to the target class.

Our pipeline consists of two main components: forward primer design and reverse primer design. For forward primer design, we first perform preprocessing to collect and validate genomic sequences and to improve data consistency. Specifically, we remove sequences that are incomplete, as well as outliers whose lengths deviate substantially from the dataset distribution (e.g., sequences that are much shorter than the typical length, such as below two-thirds of the mean, or that differ from the mean by more than one-third). We then train a C-VAE model to distinguish the target class from non-target sequences (either other variants of the same virus or sequences from other organisms). To obtain candidate primer regions, we leverage patterns captured by the convolutional layers in the encoder and extract variable-length motifs within the typical primer length range (18--25~bp). These candidates are then filtered using standard thermodynamic criteria and dimer/hairpin checks to identify viable forward primers.

Reverse primer design follows a similar procedure, with one key difference: we use the selected forward primer to locate the downstream target region from which the reverse primer should be designed. These downstream sequences, together with a synthetically generated background dataset that matches their nucleotide composition, are used as inputs to a second C-VAE model. Candidate reverse primers are obtained and filtered using the same thermodynamic and dimer-screening steps, and are paired with the forward primers to form complete primer sets. We further validate primer pairs using Primer-BLAST \cite{ye2012PrimerBLASTToolDesign} to reduce off-target amplification and to avoid primer annealing to human genomic sequences as well as closely related microbial genomes. Finally, \textit{in silico} PCR \cite{lexa2001VirtualPCR} is used to evaluate the specificity and effectiveness of the resulting primer pairs for detecting the target variant(s) or organism(s).

To our knowledge, Primer C-VAE is among the first approaches to employ a VAE-based deep-learning framework for primer design while explicitly supporting variable-length primer candidates and generating both forward and reverse primers within a unified workflow. The method is effective for sequence sub-classification tasks such as discriminating SARS-CoV-2 variants, as well as for separating organisms with large and highly similar genomes, including \textit{E. coli} and \textit{S. flexneri}. Across our experiments, Primer C-VAE achieves strong classification performance and produces highly specific primer pairs. For SARS-CoV-2 variant classification, the model reaches over 98\% accuracy, and the generated primer pairs show high specificity: they appear in more than 95\% of sequences from the target variant while occurring in less than 5\% of sequences from other variants (with Omicron as an exception, where the corresponding values are approximately 80\% and 20\%). For \textit{E. coli} and \textit{S. flexneri}, the method achieves over 96\% accuracy, and the resulting primers exhibit comparable specificity (above 95\% in target sequences and below 5\% in non-target sequences). A practical advantage of the designed primer pairs for SARS-CoV-2 is that they yield short amplicons (typically $<200$~bp), which facilitates downstream assay development across multiple PCR modalities, including conventional PCR, RT-PCR, and qPCR. Overall, Primer C-VAE provides a semi-automated workflow that reduces manual screening effort, scales to long genomic sequences, and improves the reliability of primer design for rapidly evolving pathogens and closely related organisms.

\subsection{Related work}

Recent advances in computational biology have enabled a wide range of computational strategies for primer design in molecular diagnostics. Primer3, first released in 2000, remains one of the most widely used open-source tools for primer design \cite{koressaar2007EnhancementsModificationsPrimer, untergasser2012Primer3newCapabilitiesInterfaces}. It implements established thermodynamic models to estimate oligonucleotide melting temperatures during hybridisation and to assess the stability of potential secondary structures. Although Primer3 is effective for many species-level detection tasks in viral and bacterial assays, it often produces a large set of candidates that still require downstream screening, and it has limited capability for discriminating among closely related variants within the same viral species (e.g., SARS-CoV-2 lineages). In addition, its maximum input length of 10{,}000~bp restricts direct application to long genomes, including bacterial pathogens such as \textit{Escherichia coli} with genomes on the order of 4.5--5.5~Mb.

To address these limitations, several alternative approaches have been proposed. One strategy uses finite state machines (FSMs) to classify primers into suitable and unsuitable categories \cite{ashlock2004TrainingFiniteState}. This method can complement Primer3 by prioritising higher-quality candidates from its output. However, it typically requires a sufficiently large and representative primer training set, which can limit applicability to rapidly evolving pathogens such as SARS-CoV-2, where frequent updates or retraining may be necessary.

Genetic algorithm-based approaches provide another direction for primer design \cite{wu2004PrimerDesignUsing}. In principle, these methods can overcome fixed input-length constraints and may improve the search for feasible primers compared with rule-based generation. Nevertheless, they have several practical limitations: reliance on stochastic mutation and crossover can lead to instability when mutation patterns are complex or rapidly changing; species- or dataset-specific tuning is often required for hyperparameters such as crossover probability ($p_c$), mutation probability ($p_m$), and population size ($p$); and, like many heuristic optimisation methods, they may converge to local optima rather than globally optimal solutions. Moreover, while such approaches have been used primarily for species-level primer design, they generally do not directly support robust variant-specific detection.

More recent evolutionary algorithm methods have been developed to design both forward and reverse primers for SARS-CoV-2 variant detection \cite{rincon2021DesignSpecificPrimer}. A common characteristic of these approaches is their reliance on pre-identified mutation regions (or ``signature'' mutations) to guide primer discovery. This dependence can reduce robustness when mutation hotspots shift over time and may limit generalisability to other pathogens when variant-defining mutations are not well characterised. In settings where mutation information is unavailable or incomplete, these methods may be difficult to apply.

Machine learning approaches, such as those in \cite{lopez-rincon2021ClassificationSpecificPrimer}, further improve automation by enabling primer generation without explicit reliance on known mutation sites and without strict input-length constraints. Compared with genetic and evolutionary algorithms, these methods can be computationally efficient and more stable across datasets. However, important limitations remain: existing implementations typically generate only forward primers of a fixed length (e.g., 21~bp), still require manual selection of reverse primers based on domain expertise, and often focus on species-level discrimination rather than systematic variant-level classification.

Building on prior learning-based frameworks \cite{lopez-rincon2021ClassificationSpecificPrimer}, our method is designed to reduce manual effort in identifying discriminative genomic regions within large sequence collections. It supports long input sequences, enabling analysis of both the \textasciitilde30~kb SARS-CoV-2 genome and bacterial genomes exceeding 5~Mb, such as those of \textit{E. coli}. As summarised in Table~\ref{tab: Primer3_vs_C-VAE}, our approach improves the efficiency of primer design for sequence sub-classification compared with Primer3. A key component of the proposed pipeline is a VAE-based deep learning architecture that supports reverse primer design and enables variable-length primer candidates (18--25~bp) using only the target organism's genomic sequences as input. This design reduces reliance on mutation annotations and extensive preprocessing, which is particularly useful for rapidly evolving pathogens and microorganisms with large genomes, as well as in scenarios where mutation information is inaccessible or incomplete.

\subsection{Outline of the paper}

The remainder of this paper is organised as follows. Section~\ref{section_primer_design_method} describes the proposed Primer C-VAE pipeline and its four sequential computational stages. We begin with genomic data acquisition and bioinformatic preprocessing to construct sequence alignment matrices suitable for neural network input. We then present the convolutional variational auto-encoder architecture used to generate forward-primer candidates, introduce the reverse-primer design procedure, and conclude with validation steps based on BLAST sequence similarity analysis and \textit{in silico} PCR simulation.

Section~\ref{sec:Numerical Experiment and Results} evaluates the method in two applications. In Section~\ref{sec:SARS-CoV-2 emerging variant primer design}, we demonstrate variant-specific primer design for SARS-CoV-2 and assess discrimination among the Alpha, Beta, Gamma, Delta, and Omicron variants. In Section~\ref{sec:Primer design for E.coli and S. flexneri}, we extend the evaluation to organisms with substantially larger genomes and design primers that discriminate between the closely related bacterial species \textit{Escherichia coli} and \textit{Shigella flexneri}.

Section~\ref{section_Final_discussion} discusses the results, compares Primer C-VAE with traditional primer design workflows, analyses the strengths and limitations of the framework, and outlines directions for future improvement. We conclude by summarising the contribution of Primer C-VAE as a computational approach for designing primers for target-specific detection, particularly in settings involving closely related variants and highly similar genomes.

The appendices provide supplementary material, including detailed protocols for data collection and preprocessing, feature extraction and evaluation metrics, workflow flowcharts, BLAST and \textit{in silico} PCR validation results, and comparative analyses with existing primer design tools.

\section{Process primer design with Primer C-VAE} \label{section_primer_design_method}

Our Primer C-VAE methodology comprises four sequential computational stages for an integrated primer design workflow (Figure~\ref{fig:overall_pipelines}). Stage~I covers genomic data acquisition and bioinformatic preprocessing to construct sequence alignment matrices suitable for neural network input. Stage~II generates forward-primer candidates using our convolutional variational auto-encoder architecture, extracting discriminative sequence patterns from the preprocessed data. Stage~III performs reverse primer design, using the forward-primer candidates identified in Stage~II as anchors to define downstream regions and derive complementary reverse primers that satisfy standard thermodynamic constraints. Stage~IV applies validation procedures, including BLAST sequence similarity analysis and \textit{in silico} PCR simulation, to assess primer specificity and amplification performance.

\begin{figure}[h]
  \centering
  \begin{tikzpicture}[
      scale=0.9,
      transform shape,
      box/.style={
          rectangle,
          draw=black,
          thick,
          rounded corners=3pt,
          minimum height=0.8cm,
          text centered,
          align=center,
          inner sep=4pt
      },
      arrow/.style={
          ->,
          thick,
          >=stealth
      }
  ]
  \node[align=center, font=\small\bfseries, anchor=west] at (-2,2) {Data\\Preprocessing};
  \node[align=center, font=\small\bfseries, anchor=west] at (-2,-0.1) {Forward\\Primer Design};
  \node[align=center, font=\small\bfseries, anchor=west] at (-2,-2.75) {Reverse\\Primer Design};
  \node[align=center, font=\small\bfseries, anchor=west] at (-2,-5.8) {BLAST and\\\textit{in silico} PCR};

  \node[box, fill=green!10, minimum width=2.5cm] (download) at (3,2) {Download Sequences};
  \node[box, fill=green!10, minimum width=2.5cm] (label) at (8,2) {Label Sequences};
  \node[box, fill=green!10, minimum width=2.5cm] (selection) at (13,2) {Data Selection};

  \node[box, fill=yellow!30, minimum width=2.5cm] (check1) at (2.9,0) {Forward Primer\\Frequency and\\Suitability Checks};
  \node[box, fill=cyan!10, minimum width=2.5cm] (generate1) at (8,0) {Generate\\Candidate Primers};
  \node[box, fill=orange!10, minimum width=2.5cm] (train1) at (13,0) {Train C-VAE Model};

  \node[box, fill=cyan!10, minimum width=2cm] (forward) at (2.9,-2) {Forward Primers};
  \node[box, fill=green!10, minimum width=2cm] (target) at (2.9,-3.5) {Target\\Sequences};
  \node[box, fill=green!10, minimum width=2cm] (downstream) at (6.75,-2.5) {Downstream\\Sequences};
  \node[box, fill=orange!10, minimum width=2cm] (train2) at (10.5,-2.5) {Train C-VAE Model};
  \node[box, fill=cyan!10, minimum width=2cm] (generate2) at (14.5,-1.5) {Generate\\Candidate Primers};
  \node[box, fill=yellow!30, minimum width=2cm] (check2) at (14.5,-3.5) {Reverse Primer\\Frequency and\\Suitability Checks};

  \node[box, fill=yellow!30, minimum width=2.5cm] (blast) at (2.9,-6) {Primer-BLAST and\\\textit{in silico} PCR};
  \node[box, fill=yellow!30, minimum width=2.5cm] (suitability) at (6.5,-6) {Primer-pair\\Suitability Checks};
  \node[box, fill=cyan!10, minimum width=2.5cm] (primers) at (10.5,- 6) {Primer Pairs};
  \node[box, fill=cyan!10, minimum width=2.5cm] (rprimers) at (10.5,- 4.5) {Reverse Primers};
  \node[box, fill=cyan!10, minimum width=2.5cm] (fprimers) at (14,- 6) {Forward Primers};

  \draw[arrow] (download) -- (label);
  \draw[arrow] (label) -- (selection);
  \draw[arrow] (selection) -- (train1);

  \draw[arrow] (train1) -- (generate1);
  \draw[arrow] (generate1) -- (check1);
  \draw[arrow] (check1) -- (forward);

  \draw[arrow] (forward) -- (target);
  \draw[arrow] (target) -| (downstream);
  \draw[arrow] (downstream) -- (train2);
  \draw[arrow] (train2) |- (generate2);
  \draw[arrow] (generate2) -- (check2);

  \draw[arrow] (check2) -| (rprimers);
  \draw[arrow] (rprimers) -- (primers);
  \draw[arrow] (fprimers) -- (primers);
  \draw[arrow] (primers) -- (suitability);
  \draw[arrow] (suitability) -- (blast);

  \end{tikzpicture}
  \caption{Primer C-VAE primer design workflow.}
  \label{fig:overall_pipelines}

  {\raggedright
  The Primer C-VAE workflow comprises four stages. \textbf{Stage I (Data acquisition and preprocessing)} downloads sequences, assigns labels, and curates datasets for model training. \textbf{Stage II (Forward primer design)} trains a C-VAE model, generates candidate forward primers, and filters them using frequency-based screening and thermodynamic suitability checks. \textbf{Stage III (Reverse primer design)} extracts downstream regions anchored by selected forward primers, trains a second C-VAE model, generates candidate reverse primers, and applies the same filtering criteria. \textbf{Stage IV (Validation)} forms primer pairs, evaluates amplicon size and primer--dimer risk, and assesses specificity using Primer-BLAST followed by \textit{in silico} PCR simulation.\par}
\end{figure}

The following sections provide an overview of our computational methodology. We begin with genomic data acquisition and bioinformatic preprocessing, and then detail the primer design framework, including neural-network-based feature extraction and the quantitative metrics used for selecting forward and reverse primer candidates. We conclude by describing the primer-pair validation procedures.

\subsection{Stage I: Data Acquisition and Pre-processing} \label{sec:Data acquisition and Pre-processing}

\paragraph{Data acquisition.}
Genomic sequence data used in this study were obtained from two established repositories: GISAID (\href{https://www.gisaid.org}{Global Initiative on Sharing Avian Influenza Data}, \cite{shu2017GISAIDGlobalInitiative}) and NCBI (\href{https://www.ncbi.nlm.nih.gov}{National Center for Biotechnology Information}, \cite{NCBI1988}). These repositories provided SARS-CoV-2 variant genomes and genomic sequences for \textit{Escherichia coli} and \textit{Shigella flexneri}. The sequences were used for multiple computational objectives: training the convolutional variational auto-encoder (C-VAE) for target classification, extracting candidate primer features from the trained network, identifying downstream regions for reverse primer design, and performing specificity checks through comparative sequence analyses. Our dataset includes 8{,}939 \textit{E. coli} and 5{,}373 \textit{S. flexneri} sequences from NCBI, together with 473{,}645 SARS-CoV-2 sequences from GISAID spanning five variants (Alpha, Beta, Gamma, Delta, and Omicron). In total, we analysed 610,000 complete SARS-CoV-2 genomes from GISAID and NCBI; the 473,645 sequences mentioned above refer to the GISAID subset used for model development, while the remaining NCBI sequences were used primarily for independent testing and appearance-rate evaluation. Variant labels follow World Health Organization (WHO) nomenclature, the Pango lineage classification system, and GISAID clade designations. Detailed sample distributions and labels are provided in Appendix Table~\ref{tab:SARS-COV-2_samples}.

\paragraph{Data pre-processing.}
Our preprocessing pipeline applies organism-specific filtering to improve length consistency. For most organisms, we compute the mean sequence length and remove sequences whose lengths deviate by more than one-third above or below the mean. For viruses with relatively compact and near-complete genomes, such as SARS-CoV-2, we use a less stringent rule and remove only sequences shorter than two-thirds of the mean length to preserve coverage. After filtering, we compute the maximum sequence length ($max\_vector$) within each dataset and standardize all sequences by padding with the nucleotide ``N'' to obtain a fixed-length representation, yielding $1 \times 1 \times max\_vector$ matrices. In the SARS-CoV-2 dataset, sequences averaged 29--30~kb with a maximum length of 31{,}079~bp; we removed only sequences shorter than 20~kb and represented each sequence as a $1 \times 1 \times 31{,}079$ matrix for neural network input. Variant-specific average lengths are reported in Appendix Table~\ref{tab:SARS-CoV-2_seq_length}.

To serve as neural network input, each standardized sequence must be converted from categorical nucleotides (A, T, C, G, and N) into numerical values \cite{zhang2021DeepLearningModel}. While common bioinformatic encodings include one-hot and $k$-mer representations, we use ordinal encoding as the default scheme (Equation~\ref{eq:ordinal_encoding}). This choice is computationally efficient and simple to implement for convolutional architectures, and has been reported to provide competitive performance in similar settings \cite{choong2017EvaluationConvolutionaryNeural}. One-hot encoding is also supported as an alternative when appropriate.

Before model training, each standardized sequence is ordinal-encoded using Equation~\ref{eq:ordinal_encoding}; An example of this transformation is shown in Appendix Figure~\ref{fig:sequence_encoding_example}.

\begin{equation} \label{eq:ordinal_encoding}
Y := f(x) := \left \{
\begin{aligned}
    0 & \quad \mbox{ if } & x=N,\\
    1 & \quad \mbox{ if } & x=C,\\
    2 & \quad \mbox{ if } & x=T,\\
    3 & \quad \mbox{ if } & x=G,\\
    4 & \quad \mbox{ if } & x=A.\\
\end{aligned}
\right.
\end{equation}

For downstream interpretation, we also implement an inverse transformation that converts numerical outputs back to nucleotide symbols. For classification tasks, taxonomic labels are encoded as integer class indices $y \in \{0,1,\ldots,C-1\}$ for model training. For example, SARS-CoV-2 variant classification uses $C=5$ classes (Alpha, Beta, Gamma, Delta, and Omicron), and each sequence is assigned a single class index. When needed for reporting or downstream analysis (e.g., confusion matrices or plots), these indices can be converted to one-hot vectors of dimension $1 \times 5$.

\paragraph{Data selection.}
After preprocessing, we apply a structured data partitioning strategy to support three computational requirements: model training, validation, and testing. During primer design, we additionally select target-class sequences to generate candidate primers and construct reference sequence sets for specificity evaluation. These reference sets support: (i) assessing primer specificity to the target class, (ii) estimating appearance frequencies to reduce off-target amplification risk, and (iii) checking non-complementarity to human genomic sequences and closely related microbial genomes to mitigate cross-reactivity. Accordingly, our data selection procedure includes three steps: (1) splitting data for training/validation/testing; (2) selecting sequences for primer candidate generation; and (3) compiling reference datasets for specificity assessment. A worked example for SARS-CoV-2 variant-specific primer design is provided in Table~\ref{tab:SARS-CoV-2_data_selection}.

\subsection{Stage II: Forward Primer Design} \label{sec:Forward primer design}

\paragraph{Primer C-VAE architecture.}
We use a dataset of pre-processed genome sequences represented as fixed-length input matrices,
$\mathcal{D} = \{seq_1, seq_2, \ldots, seq_n\}$, where each $seq_i \in \mathbb{R}^{1 \times 1 \times max\_vector}$ denotes an encoded genome sequence. The C-VAE model is trained for supervised classification with $C$ classes, where $C$ depends on the application (e.g., $C=5$ for SARS-CoV-2 variants: Alpha, Beta, Gamma, Delta, and Omicron; and $C=2$ for \textit{E. coli} versus \textit{S. flexneri}). The model learns latent representations of input sequences, which are used both for sequence reconstruction via the decoder and for class prediction via the classifier head. Figure~\ref{fig:Primer_C-VAE_architecture} illustrates the architecture.

\begin{figure}[h]
  \centering
  \begin{tikzpicture}[
      box/.style={
          rectangle,
          draw=black,
          thick,
          minimum width=2cm,
          minimum height=0.8cm,
          text centered,
          align=center,
          inner sep=4pt
      },
      arrow/.style={
          ->,
          thick,
          >=stealth
      },
      group/.style={
          rectangle,
          draw,
          dashed,
          inner sep=10pt
      }
  ]
  \node[box, fill=green!10] (input) at (4cm,-7cm) {Genomic Sequence Input};
  
  \begin{scope}[local bounding box=encoder, xshift=4cm, yshift=-2.7cm]
      \node[box, fill=orange!10] (conv1) at (0,-2.4cm) {Conv2D + ReLU};
      \node[box, fill=orange!10, above=0.6cm of conv1] (mp1) {MaxPool2D};
      \node[box, fill=orange!10, above=0.6cm of mp1] (conv2) {Conv2D + ReLU};
      \node[box, fill=orange!10, above=0.6cm of conv2] (mp2) {MaxPool2D};
      \node[box, fill=blue!10, above=0.6cm of mp2] (fc) {FC + ReLU + BN};
  \end{scope}
  
  \begin{scope}[local bounding box=latent, xshift=8cm]
      \node[box, fill=yellow!30] (mu) at (0,0.6cm) {$\mu$};
      \node[box, fill=yellow!30] (logvar) at (0,-0.6cm) {$\log\sigma^2$};
      \node[box, fill=yellow!40, right=3cm of $(mu)$] (z) {z};
  \end{scope}
  
  \begin{scope}[local bounding box=classifier, xshift=9.5cm, yshift=-3cm]
      \node[box, fill=cyan!10] (cl1) {FC + ReLU + BN};
      \node[box, fill=cyan!10, below=0.6cm of cl1] (cl2) {FC + ReLU + BN};
      \node[box, fill=cyan!10, below=0.6cm of cl2] (cl3) {FC + ReLU + BN};
      \node[box, fill=cyan!10, below=0.6cm of cl3] (clout) {Output Layer};
  \end{scope}
  
  \begin{scope}[local bounding box=decoder, xshift=16cm, yshift=0.6cm]
      \node[box, fill=blue!10] (fcd) {FC Decoder};
      \node[box, fill=orange!10, below=0.6cm of fcd] (ct1) {ConvTranspose2D};
      \node[box, fill=orange!10, below=0.6cm of ct1] (ct2) {ConvTranspose2D};
  \end{scope}
  
  \node[box, fill=gray!20, right=1.5cm of clout] (output2) {Classification Labels};
  \node[box, fill=green!10, below=1cm of ct2] (output1) {Reconstructed Sequence};
  
  \node[group, fit=(conv1) (mp1) (conv2) (mp2) (fc), label=above:Encoder] {};
  \node[group, fit=(mu) (logvar) (z), label=above:Latent Space] (latent_box) {};
  \node[group, fit=(cl1) (cl2) (cl3) (clout), label=above:Classifier] {};
  \node[group, fit=(fcd) (ct1) (ct2), label=above:Decoder] {};
  
  \draw[arrow] (conv1) -- (mp1);
  \draw[arrow] (mp1) -- (conv2);
  \draw[arrow] (conv2) -- (mp2);
  \draw[arrow] (mp2) -- (fc);
  
  \draw[arrow] (input) |- ($(conv1.south)+(0,-0.3)$) -- (conv1);
  
  \coordinate (entry) at ($(latent_box.west)+(-0.5,0.0cm)$);
  \draw[arrow] (fc) -| (entry) |- (mu);
  \draw[arrow] ($(fc)!1.0!(entry)$) |- (logvar);
  
  \draw[arrow] (mu.east) -- node[below, yshift=-24pt, xshift=12pt] {perturbation} (z.west);
  \draw[arrow] (logvar.east) -- node[below, yshift=-16pt, xshift=12pt] {function} (z.west);
  
  \draw[arrow] (z) |- (cl1);
  \draw[arrow] (z) -- (fcd);
  
  \draw[arrow] (cl1) -- (cl2);
  \draw[arrow] (cl2) -- (cl3);
  \draw[arrow] (cl3) -- (clout);
  
  \draw[arrow] (fcd) -- (ct1);
  \draw[arrow] (ct1) -- (ct2);
  
  \draw[arrow] (clout) -- (output2);
  \draw[arrow] (ct2) -- (output1);
  
  \end{tikzpicture}
  \caption{Primer C-VAE architecture}
  \label{fig:Primer_C-VAE_architecture}

  {\raggedright \textbf{Note:} Primer C-VAE consists of three components: (1) a convolutional encoder that extracts hierarchical features from genomic inputs, (2) a variational latent space in which a latent vector $z$ is sampled using the reparameterization trick from the learned parameters $\mu$ and $\log\sigma^2$, and (3) two output heads---a classifier for sequence categorization and a decoder for sequence reconstruction. The model is trained end-to-end by jointly optimizing the classification objective and the reconstruction objective, encouraging the latent representation to be both discriminative and informative. \par}

\end{figure}

The C-VAE model uses a five-layer encoder: two convolutional layers (Conv2D) with ReLU activations, each followed by a max-pooling layer (MaxPool2D), and a final fully connected (FC) layer with ReLU and batch normalization (BN). This convolutional hierarchy progressively extracts sequence features, where earlier layers capture short, motif-like patterns and deeper layers capture higher-level combinations of these patterns. ReLU activations improve optimization behavior and mitigate vanishing gradients during backpropagation \cite{nairRectifiedLinearUnits}. Max-pooling reduces the effective dimensionality while retaining salient features, enabling the FC layer to aggregate multi-scale information for downstream tasks.

Within the variational component, the latent representation $z$ is obtained via the reparameterization trick, using encoder outputs $\mu$ and $\log\sigma^2$ to define a Gaussian distribution from which $z$ is sampled. From $z$, the architecture branches into two pathways: a classification head that predicts sequence labels, and a decoder that reconstructs the input using an approximately inverted architecture.

The decoder serves two roles in our pipeline. First, reconstruction provides a regularization signal that encourages the latent representation to preserve information about the input sequence. Second, by comparing reconstructed sequences with the original inputs, we can highlight regions where reconstruction errors are consistently higher across classes, which can indicate genomic loci with increased variability between variants. Although such regions are not guaranteed to be the sole drivers of classification performance, they can provide useful biological cues for primer design by prioritizing candidate segments that differ between target and non-target groups. This dual-objective design therefore supports both discriminative learning and downstream interpretation.

\subsubsection{Model Training}

During preprocessing, input genomic sequences are standardized to a uniform length using the maximum sequence length within each dataset as the reference. Each sequence is represented as a $1 \times 1 \times max\_vector$ tensor to ensure compatibility with the C-VAE architecture. After data curation, the C-VAE model is trained to learn discriminative genomic features that separate target classes from non-target classes. In the SARS-CoV-2 experiment, this is formulated as a five-class classification task, where each sequence is assigned to one of $$\mathcal{C}=\{ \text{Alpha, Beta, Gamma, Delta, Omicron} \}$$

Training uses the Adam optimizer \cite{kingma2017AdamMethodStochastic} with an adaptive learning rate, together with a multi-class classification objective ($\mathcal{L}_{class}$). Specifically, the classifier head outputs logits $s \in \mathbb{R}^{5}$, and $\mathcal{L}_{class}$ is defined as the categorical cross-entropy loss:
$$
\mathcal{L}_{class} = - \sum_{k=1}^{5} y_k \log p_k, \quad \text{where } p = \mathrm{softmax}(s).
$$
In practice, $\mathcal{L}_{class}$ is implemented using PyTorch's \texttt{CrossEntropyLoss}, which takes logits as input and applies the softmax operation internally for numerical stability \cite{paszke2019PyTorchImperativeStyle}.

The training objective combines multiple loss terms. The reconstruction loss ($\mathcal{L}_{recon}$), defined as the mean squared error between the input sequence and its reconstruction, encourages the latent representation to preserve information about the input. In addition, the Kullback--Leibler divergence loss ($\mathcal{L}_{KL}$) acts as a regularizer that encourages the approximate posterior to remain close to a standard normal prior, given by:
$$
\mathcal{L}_{KL} = -0.5\sum(1 + \log\sigma^2 - \mu^2 - \sigma^2).
$$

The overall loss function is:
$$
\mathcal{L}_{total} = \mathcal{L}_{recon} + \beta \cdot \mathcal{L}_{KL} + \lambda_{class} \cdot \mathcal{L}_{class} + \lambda_{reg} \cdot \mathcal{L}_{reg}.
$$

Here, $\mathcal{L}_{reg}$ denotes an L2 regularization term that mitigates overfitting by penalizing large parameter magnitudes. The hyperparameters $\beta$, $\lambda_{class}$, and $\lambda_{reg}$ serve as weighting coefficients that control the relative contributions of each term. This combined objective encourages the model to learn latent representations that are both informative for reconstruction and discriminative for variant classification, which supports downstream primer candidate identification.

\subsubsection{Feature Extraction and Forward Primer Generation}

After training the C-VAE model, we extract candidate forward primers using four feature-identification strategies. Three strategies derive candidate regions from activation patterns in the convolutional encoder, whereas the fourth leverages differences between the input sequence and the decoder reconstruction.

The first convolutional layer in the encoder is designed to capture short, motif-like patterns and consists of 12 filters with a $1 \times N$ kernel, where $N$ is set to match the desired primer length. In practice, we generate candidate regions within the typical primer range (18--25~nt) by applying post-processing to the activation maps of this first convolutional layer using three encoder-based methods: (1) Pooling, (2) Top-$k$, and (3) Mix. In addition, a decoder-based method, (4) Reconstruction, identifies informative positions by analyzing discrepancies in the reconstructed sequence.

\begin{figure}[htbp]
\centering
\includegraphics[width=0.6 \textwidth]{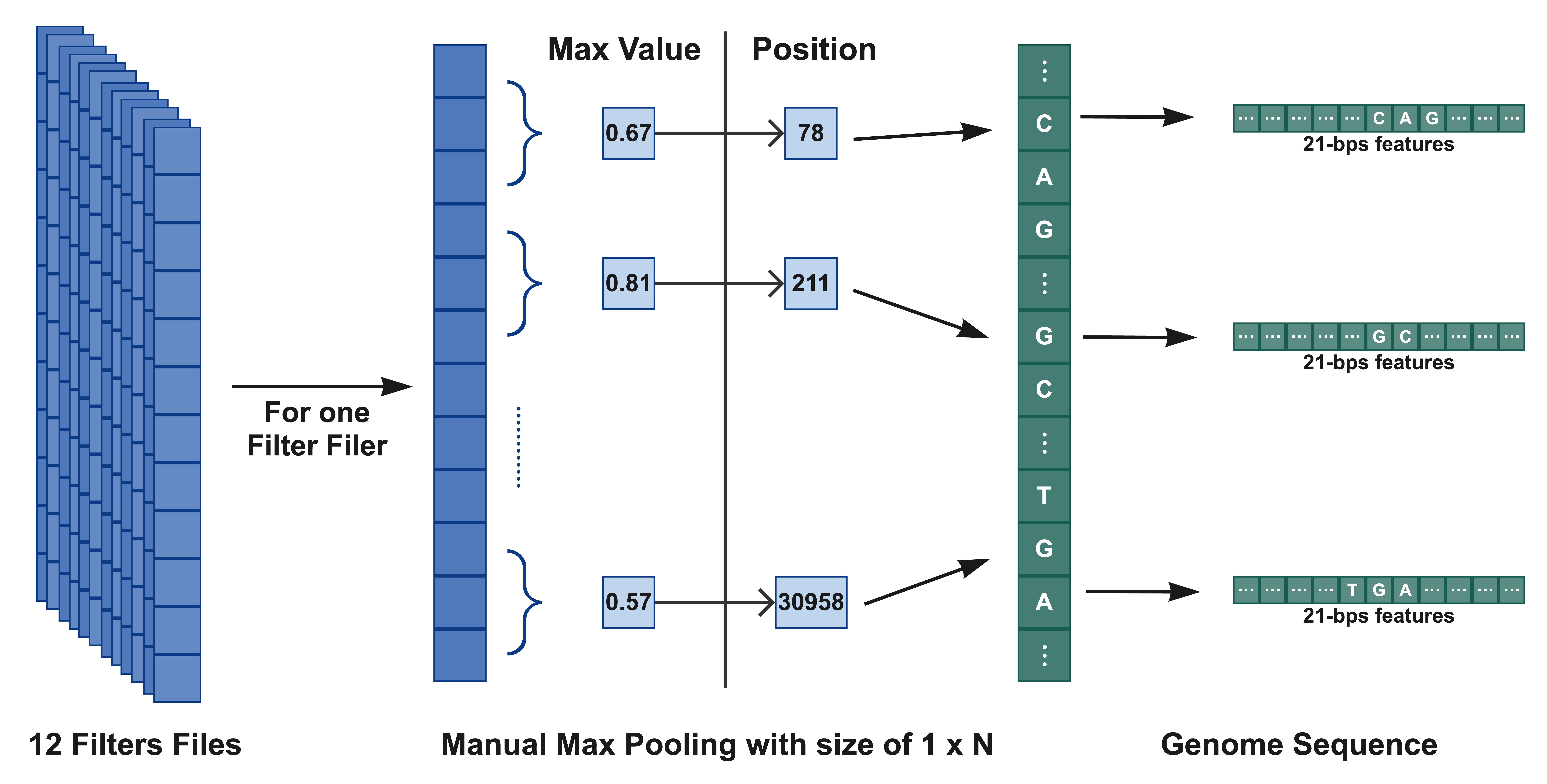}
\caption{\label{fig:Filter_maxPool}Feature extraction from filter activation maps by simulating max-pooling with index tracking.}
\end{figure}

\paragraph{Pooling method.}
This method is inspired by max-pooling, but with an important modification: instead of retaining only the maximum value, we record the \emph{position} of the maximum activation within each pooling window. Concretely, for each filter activation map, we apply a custom max-pooling operation that stores argmax indices (Figure~\ref{fig:Filter_maxPool}). The nucleotide positions corresponding to these maxima are written to a position file. This procedure is applied independently to all 12 filters to capture a diverse set of candidate positions.

\paragraph{Top-$k$ method.}
Unlike pooling, the Top-$k$ method performs global selection without predefined pooling windows. Given a user-specified parameter $k$, we identify the $k$ largest activation values across the full activation map and record their positions. We selected $k$ via empirical analysis on the SARS-CoV-2 Alpha and Delta datasets, evaluating $k \in \{75, 125, 175, 250\}$, corresponding to approximately 0.25\%, 0.50\%, 0.58\%, and 0.83\% of a \textasciitilde30{,}000-nt genome. As reported in Appendix Table~\ref{tab:Top_method}, $k=175$ (0.58\%) provided a good trade-off between computational cost and primer generation performance, while maintaining high appearance rate within the target variant.

\paragraph{Mix method.}
The Mix method combines the Pooling and Top-$k$ strategies. We first apply pooling to partition the activation map into windows, and then record the top-$k$ positions within each window. Based on empirical testing, we use a pooling window size of $1 \times 500$ and record the top 10 positions per window.

\begin{figure}[H]
  \centering
  \includegraphics[width=0.6 \textwidth]{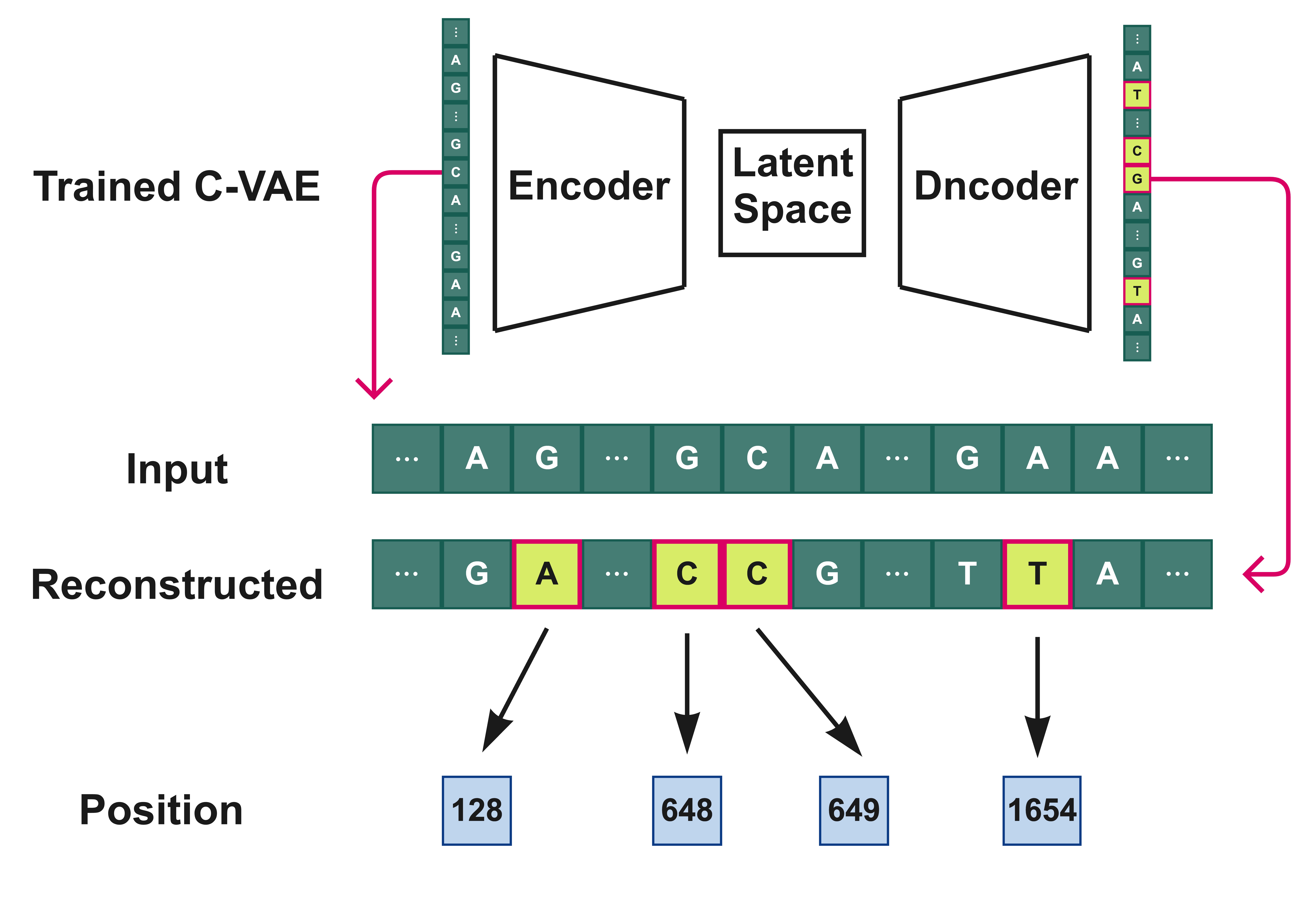}
  \caption{\label{fig:Reconstruction_method}Feature extraction from reconstructed sequences by identifying nucleotide divergence.}
\end{figure}

\paragraph{Reconstruction method.}
This method uses the decoder reconstruction to identify candidate positions. Although the reconstructed sequence has the same length as the input, it may differ at specific nucleotide positions. Rather than treating these discrepancies as errors, we use them as a signal: positions with consistently higher reconstruction divergence can indicate regions that are less well captured by the shared representation or that differ systematically between classes. Such regions may be informative for primer design, for example by highlighting loci that vary between target and non-target groups or loci with elevated variability within variant populations. We therefore record positions with the highest divergence between the input and the reconstructed sequence (Figure~\ref{fig:Reconstruction_method}).

Our forward-primer design procedure begins by using the extracted positions as anchors for primer construction. For each selected position, we retrieve the corresponding nucleotide in the input sequence and extend to include flanking bases to reach a user-defined primer length $L \in [18,25]$~nt. For example, when $L=25$, we construct a 25-nt primer by taking the anchor nucleotide plus 12~nt upstream and 12~nt downstream. For even lengths, we extend asymmetrically using $\lfloor(L-1)/2\rfloor$ bases on one side and $\lceil(L-1)/2\rceil$ on the other to obtain exactly $L$ nucleotides.

Each candidate primer is then filtered using standard quality criteria, including thermodynamic properties, dimer/hairpin risk, and appearance rate in the target dataset. As shown in Figure~\ref{fig:feature_extraction_flow}, candidates must satisfy established primer design constraints \cite{dieffenbach1993GeneralConceptsPCR, AddgeneProtocolHow} before being retained as viable forward primers:

\begin{enumerate}
    \item Length of 18--25~nt;
    \item GC content of 40--60\%;
    \item 1--2 G/C bases at both the 5$'$ and 3$'$ ends (GC clamp);
    \item Melting temperature (Tm) of 45--60$^{\circ}$C (calculated with Primer3; see also \cite{MeltingTemperatureTm});
    \item For downstream pairing, forward and reverse primers should have a Tm difference within 5$^{\circ}$C;
    \item No strong self-complementarity (e.g., no more than five consecutive complementary bases), assessed using IDT OligoAnalyzer \cite{OligoAnalyzerToolPrimer}.
\end{enumerate}

\begin{figure}[htbp]
  \centering
  \begin{tikzpicture}[
      box/.style={
          draw,
          minimum width=1.6cm,
          minimum height=0.7cm,
          rounded corners=2pt,
          align=center,
          font=\small
      },
      method box/.style={
          draw,
          minimum width=1.8cm,
          minimum height=0.5cm,
          rounded corners=2pt,
          fill=orange!10,
          align=center,
          font=\small
      },
      arrow/.style={
          ->,
          >=latex,
          thick,
          shorten >=2pt,
          shorten <=2pt
      }
  ]
      \node[box, fill=blue!10] (input) at (-3.1,0) {Input\\Sequence};
      \node[box, fill=green!10] (encoder) at (-1,0) {Encoder};
      \node[box, fill=pink!20] (position) at (5,0) {Position\\Files};
      \node[box, fill=cyan!10] (primer) at (7.2,0) {Primer\\Generation};
      \node[box, fill=green!20] (validation) at (9.4,0) {Criteria\\Validation};
      \node[box, fill=blue!20] (final) at (11.5,0) {Forward\\Primers};

      \node[method box] (pooling) at (2,1.2) {Pooling};
      \node[method box] (top) at (2,0.4) {Top-$k$};
      \node[method box] (mix) at (2,-0.4) {Mix};
      \node[method box] (reconstruction) at (2,-1.2) {Reconstruction};

      \draw[dashed] (-0.1,-1.6) rectangle (4.1,1.6);
      \node[font=\small] at (2,1.9) {Feature Extraction};

      \coordinate (start) at (0.3,0);
      \coordinate (end) at (3.7,0);

      \draw[arrow] (input) -- (encoder);
      \draw[arrow] (position) -- (primer);
      \draw[arrow] (primer) -- (validation);
      \draw[arrow] (validation) -- (final);

      \draw[arrow] (encoder) -- (start);
      \draw[arrow] (start) |- (pooling);
      \draw[arrow] (start) |- (top);
      \draw[arrow] (start) |- (mix);
      \draw[arrow] (start) |- (reconstruction);

      \draw[arrow] (pooling) -| (end);
      \draw[arrow] (top) -| (end);
      \draw[arrow] (mix) -| (end);
      \draw[arrow] (reconstruction) -| (end);
      \draw[arrow] (end) -- (position);
  \end{tikzpicture}
  \caption{Computational workflow for feature extraction and forward primer design. Candidate positions are identified by four feature extraction methods and recorded in position files, which guide primer construction. Candidate primers are then filtered using thermodynamic and specificity criteria.}
  \label{fig:feature_extraction_flow}
\end{figure}

To assess specificity, we compute the appearance rate of each candidate primer in target and non-target sequence sets. Primers that fail to bind to the target class or that match multiple loci can reduce assay performance and increase the risk of non-specific amplification \cite{schrick2016PitfallsPCRTroubleshooting}. Unless otherwise stated, the appearance rate is computed using an \textbf{exact-match} rule (0 mismatches): a primer is counted as present in a sequence only if its nucleotide string (or its reverse complement) appears as an exact contiguous substring. This strict filter provides a conservative first-pass specificity screen; potential mismatch-tolerant binding and off-target amplification are further evaluated in Stage IV via thermodynamic analysis and \textit{in silico} PCR tools.

\subsection{Stage III: Reverse Primer Design} \label{sec:Reverse_primer_design}

In conventional PCR assay design, the reverse primer is placed downstream of the forward primer on the reference (5$' \rightarrow$3$'$) sequence, and the distance between the two primer binding sites determines the amplicon length. For example, when a short product is desired (e.g., $\sim$200~bp for qPCR), the reverse primer is selected such that the resulting amplicon length falls within the target range.

Building on this principle, we develop a deep-learning-assisted procedure for reverse primer design. Unlike forward primer generation, reverse primer design in our pipeline is conditioned on two inputs: (1) validated forward primers and (2) the corresponding target-class genome sequences. Using the forward primers as anchors, we restrict the search space to biologically relevant downstream regions, which improves efficiency and reduces unnecessary scanning of the full genome.

\subsubsection{Data pre-processing for generating the downstream dataset}

Reverse primer design requires target-class sequences and validated forward primers. We first locate each validated forward primer within each target sequence and then define the downstream region in which the reverse primer binding site must lie, i.e., the segment extending from the end of the forward primer to the 3$'$ terminus on the reference sequence. This reflects the standard convention that sequences are indexed in the 5$' \rightarrow$3$'$ direction, and reverse primers are chosen downstream of the forward primer to yield the desired amplicon length.

\begin{figure}[htbp]
    \begin{minipage}{0.55\textwidth} 
        \centering
        \includegraphics[width=\linewidth]{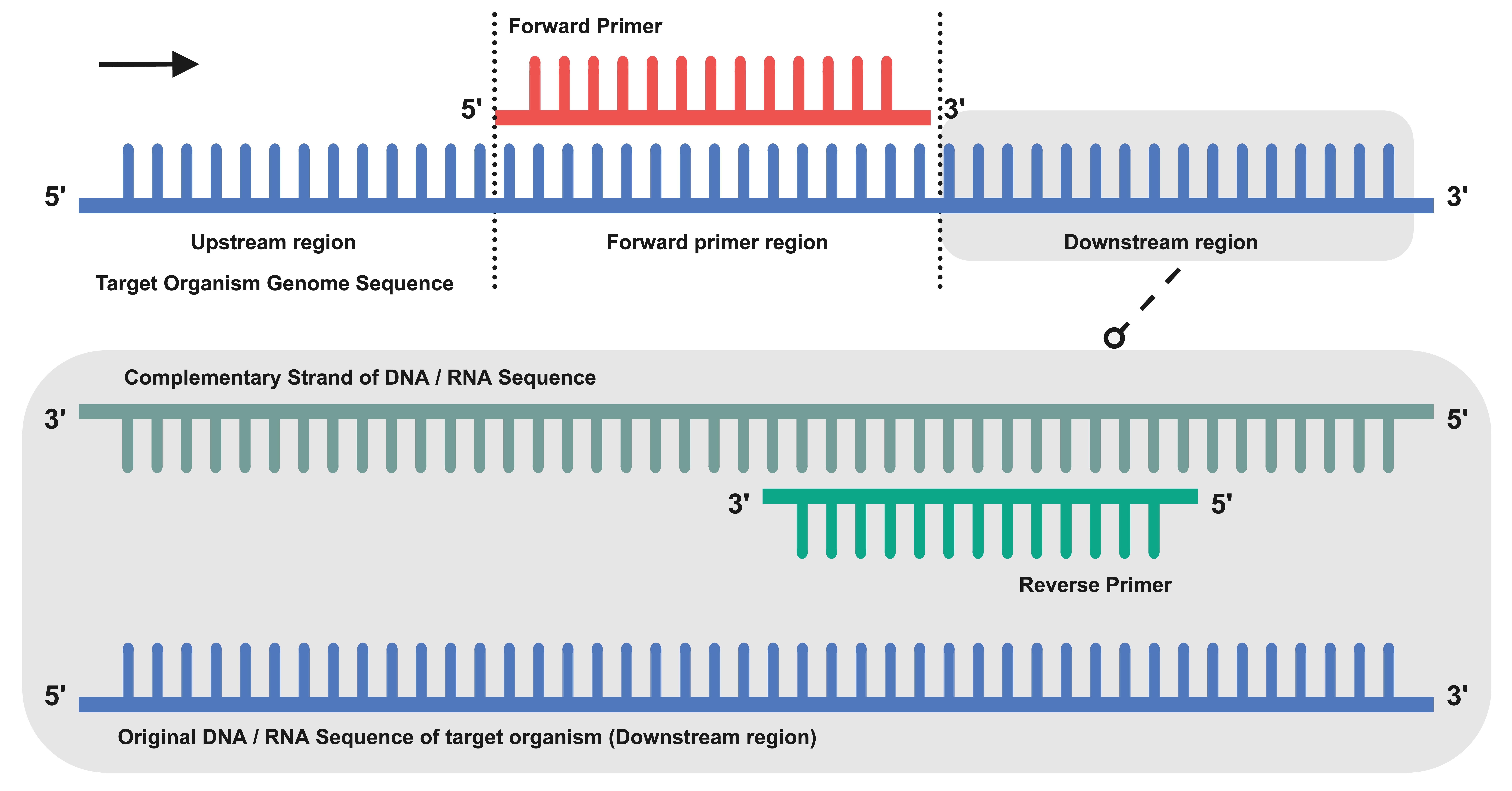}
        \caption{Functional segmentation of target sequences for reverse primer design. Each sequence is partitioned into upstream, forward primer, and downstream regions. The start position and length of the downstream region depend on the binding location of the validated forward primer in that sequence.}
        \label{fig:Downstream_region}
    \end{minipage}
    \hfill  
    \begin{minipage}{0.4\textwidth}
        \centering
        \includegraphics[width=\linewidth]{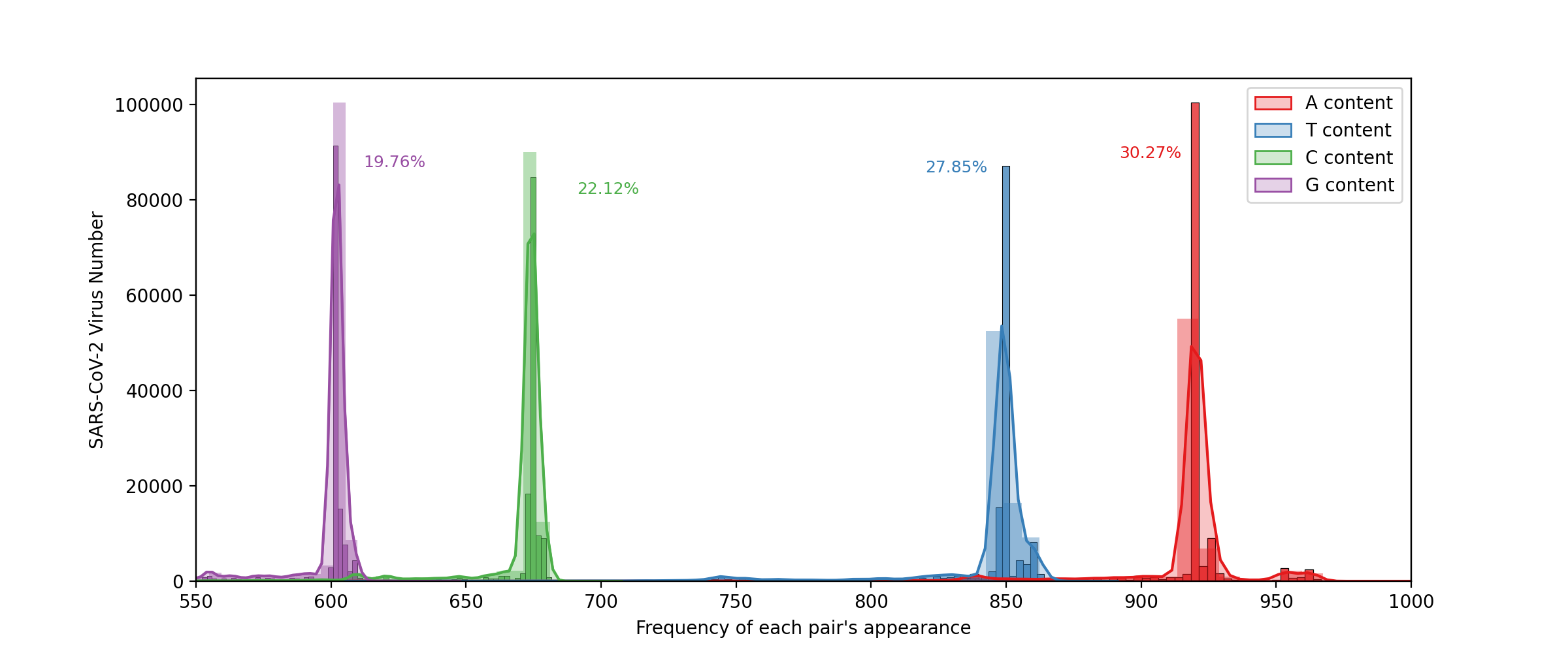}
        \caption{Nucleotide composition analysis of downstream regions in the target dataset.}
        \label{fig:ATCG_content}
    \end{minipage}
\end{figure}

For each target sequence containing a validated forward primer, we partition the sequence into three regions (Figure~\ref{fig:Downstream_region}):

\begin{enumerate}
    \item Upstream region: from the 5$'$ terminus to the nucleotide immediately preceding the forward primer;
    \item Forward primer region: the forward primer binding segment;
    \item Downstream region: from the last nucleotide of the forward primer to the 3$'$ terminus.
\end{enumerate}

For reverse primer design, we retain only the downstream region and construct a downstream dataset. This targeted preprocessing step confines the candidate search space to biologically valid locations and substantially reduces computational complexity.

\subsubsection{Generation of Synthetic Downstream Data}

We adapt the C-VAE framework used in Stage~II to reverse primer design. A key challenge is that the downstream dataset constructed for a given target variant (or target class) contains only one biological label, which provides no explicit contrast for supervised discrimination. Without a meaningful negative class, the model may learn features that satisfy the training objective without capturing class-specific biological constraints.

To introduce a contrasting label, we construct a synthetic reference dataset and formulate reverse primer learning as a binary classification task: real downstream sequences versus synthetic downstream sequences. The synthetic sequences are generated to match the nucleotide composition of the real downstream dataset. Specifically, we estimate the empirical A/T/C/G frequencies from authentic downstream regions (Figure~\ref{fig:ATCG_content}) and then sample random sequences that preserve these mononucleotide proportions. This provides a controlled negative class that is statistically similar at the nucleotide-frequency level but does not reflect biological structure shaped by evolutionary and functional constraints.

For clarity, consider the SARS-CoV-2 Delta example with the validated forward primer \texttt{CTACCGCAATGGCTTGTCTTG}. Our procedure consists of four steps: (1) select Delta sequences; (2) extract downstream regions based on the forward primer location; (3) compute nucleotide composition statistics; and (4) generate a synthetic downstream dataset that matches these statistics.

\subsubsection{Model Training and Feature Extraction}

The C-VAE architecture and loss formulation remain the same as in Stage~II, but the training data and labels differ. We train the model to discriminate authentic downstream sequences from synthetic sequences under this binary classification setup. Although synthetic sequences match the overall nucleotide composition of real downstream regions, they lack higher-order patterns and biological constraints present in genuine genomes. Consequently, the trained model can learn features that reflect biologically structured sequence patterns rather than nucleotide frequencies alone. These learned features are then used to prioritize candidate regions for reverse primer generation in the downstream dataset.

\paragraph{Reverse primer candidate extraction.}
After training on the real-versus-synthetic downstream datasets, we extract candidate reverse primers from the downstream regions using the same four feature extraction strategies described in Stage~II (Pooling, Top-$k$, Mix, and Reconstruction). Concretely, we apply the encoder-based methods to the first convolutional-layer activation maps of downstream inputs and record high-activation nucleotide positions in position files, and we apply the Reconstruction method by identifying positions with high divergence between the downstream input and its reconstruction. These positions serve as anchors for candidate construction, identical to forward primer generation: for a desired primer length $L \in [18,25]$~nt, we extend around each anchor position within the downstream sequence to obtain an $L$-nt candidate. Candidates that cannot be formed due to proximity to the downstream boundary are discarded.

Because reverse primers anneal to the complementary strand, each extracted candidate sequence is converted to its reverse complement before thermodynamic screening and pairing with the corresponding forward primer. This reuse of the Stage~II extraction procedures ensures that forward and reverse primer candidates are generated in a consistent, model-driven manner while restricting reverse primer search to biologically valid downstream regions.

\subsection{Stage IV: \textit{in silico} PCR and Primer-BLAST Validation}

Following reverse primer design, all candidate primer pairs undergo \textit{in silico} validation using standard virtual PCR simulation protocols \cite{lexa2001VirtualPCR}. This step allows us to assess expected amplification behavior and primer specificity before laboratory testing. Reverse primer candidates are generated using the same C-VAE architecture and the same four feature extraction strategies as in forward primer design. For validation, we use three complementary tools: (1) FastPCR \cite{kalendar2017FastPCRSilicoTool} for thermodynamic analysis and \textit{in silico} PCR; (2) Unipro UGENE \cite{okonechnikov2012UniproUGENEUnified} for thermodynamic analysis and \textit{in silico} PCR; and (3) Primer-BLAST \cite{ye2012PrimerBLASTToolDesign} for genome-scale specificity assessment.

We evaluate each primer pair against two criteria: (i) successful amplification of the intended target with an amplicon length within the desired size range, and (ii) no detectable off-target amplification products when tested against the selected reference genomes. Primer pairs are considered \textit{in silico} validated only if they satisfy both criteria consistently across all three tools. Candidates that pass this \textit{in silico} screening are then prioritized for downstream experimental evaluation. This multi-tool validation strategy provides an additional quality-control layer and helps reduce the number of unsuitable primer pairs carried forward to wet-lab testing. Throughout this paper, the term "validation" refers to to computational screening (does not constitute wet-lab validation) including \textit{in silico} validation via Primer-BLAST, FastPCR, and UGENE, unless explicitly stated otherwise.


\section{Numerical Experiment and Results} \label{sec:Numerical Experiment and Results}

\subsection{Experiment 1: SARS-CoV-2 Emerging Variant Primer Design} \label{sec:SARS-CoV-2 emerging variant primer design}

Although the acute public health emergency has subsided, SARS-CoV-2 continues to evolve, giving rise to new lineages that may warrant continued surveillance \cite{Wang_Chen_Hozumi_Yin_Wei_2022, Christensen_2022, He_Hong_Pan_Lu_Wei_2021}. According to World Health Organization (WHO) surveillance data, more than 700 million COVID-19 cases had been reported globally by October 2024, with reported deaths exceeding 7 million \cite{COVID-19_dashboard}. The recent detection of the XEC lineage---a recombinant derived from the KS.1.1 and KP.3.3 lineages---in multiple European countries and the United Kingdom \cite{Arora_2024} further highlights the ongoing need for robust molecular detection systems capable of rapidly identifying novel genomic signatures.

\begin{table}[H]
\centering
\begin{tabular}{c|ccc|cc}
 & Training set & Validation set & Test set & Generated primers & Calculated appearance\\ \hline
Source & GISAID & GISAID & NCBI & GISAID &  GISAID and NCBI \\ \hline
Alpha & 2,000 & 2,000 & 2,000 & 1,000 (or) & 5,000 \\
Beta & 2,000 & 2,000 & 2,000 & 1,000 (or) & 5,000 \\
Gamma & 2,000 & 2,000 & 2,000 & 1,000 (or) & 5,000 \\
Delta & 2,000 & 2,000 & 2,000 & 1,000 (or) & 5,000 \\
Omicron & 2,000 & 2,000 & 2,000 & 1,000 (or) & 5,000 \\
Other Taxa & 0 & 0 & 0 & 0 & 3,640 \\ \hline
Total Number & 10,000 & 10,000 & 10,000 & 1,000 & 28,640 \\ \hline
\end{tabular}
\caption{Data selection for the C-VAE model, generated and calculated appearance rate of forward primers.}
\label{tab:SARS-CoV-2_data_selection}
\end{table}

The study analyzed 610{,}000 complete SARS-CoV-2 genome sequences obtained from the GISAID and NCBI repositories. To ensure balanced representation across variants, we performed stratified random sampling from GISAID and selected 4{,}000 sequences per variant. These 20{,}000 GISAID sequences were then split into a training set (2{,}000 per variant; 10{,}000 total) and a validation set (2{,}000 per variant; 10{,}000 total), yielding an equal class distribution (1:1:1:1:1) in both splits (Table~\ref{tab:SARS-CoV-2_data_selection}). During training, we used a mini-batch size of 50, corresponding to 200 training batches per epoch for the 10{,}000-sequence training set, which helps accommodate memory constraints \cite{he2019ControlBatchSize} and reduces the impact of ordering effects during optimization \cite{schuster2019DebiasingFactVerification}. We additionally constructed an independent test set composed exclusively of NCBI sequences (2{,}000 per variant; 10{,}000 total), with lineage labels assigned using the PANGOLIN classification system \cite{otoole2021AssignmentEpidemiologicalLineages}.

\begin{figure}[htp]
  \begin{minipage}{0.48\textwidth}
      \centering
      \includegraphics[scale=0.47]{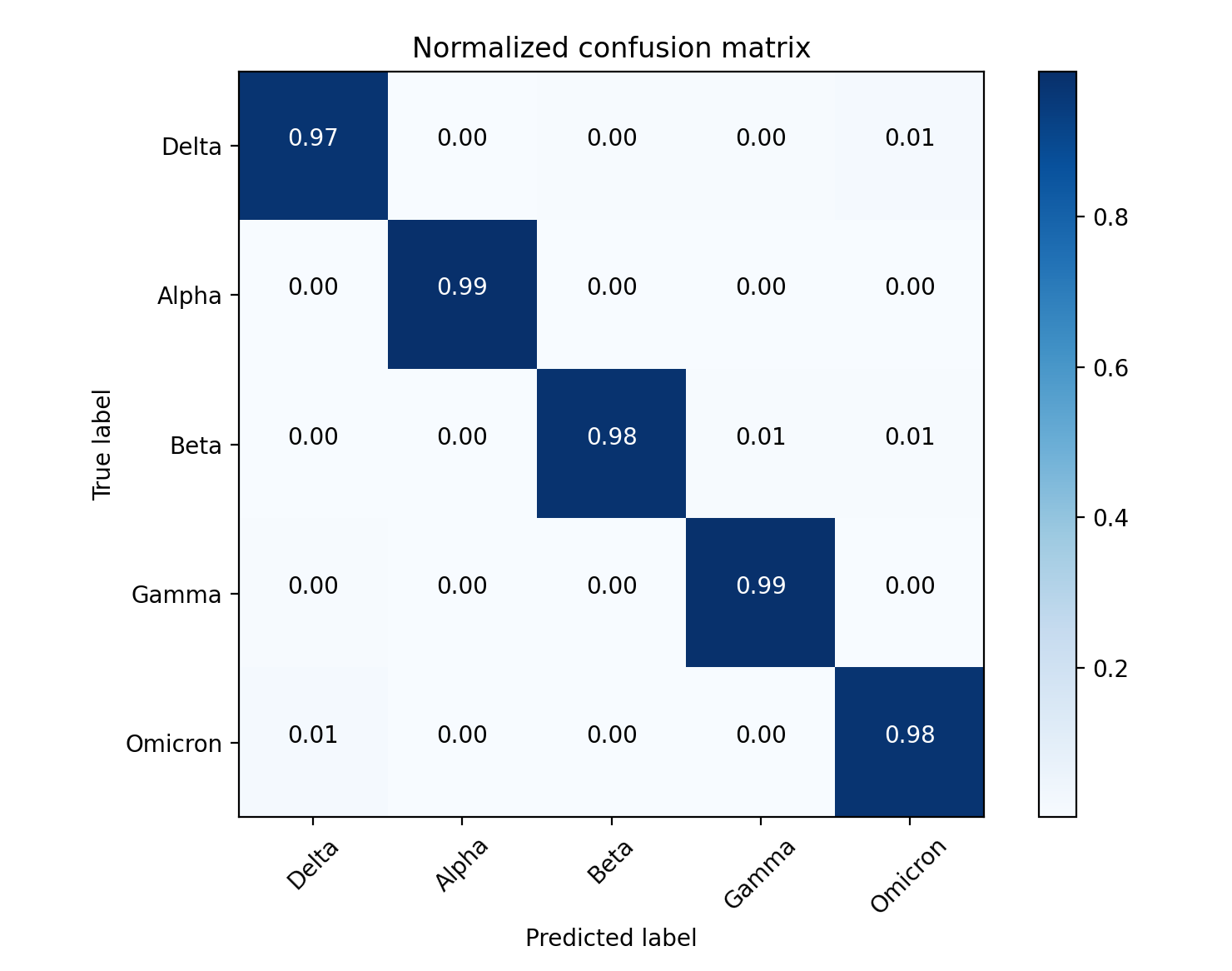}
      \caption{Confusion matrix for five-class SARS-CoV-2 variant classification using the C-VAE model (2{,}000 sequences per variant in Test set).}
      \label{fig:Confusion_Matrix}
  \end{minipage}
  \hfill
  \begin{minipage}{0.48\textwidth}
      \centering
      \includegraphics[scale=0.56]{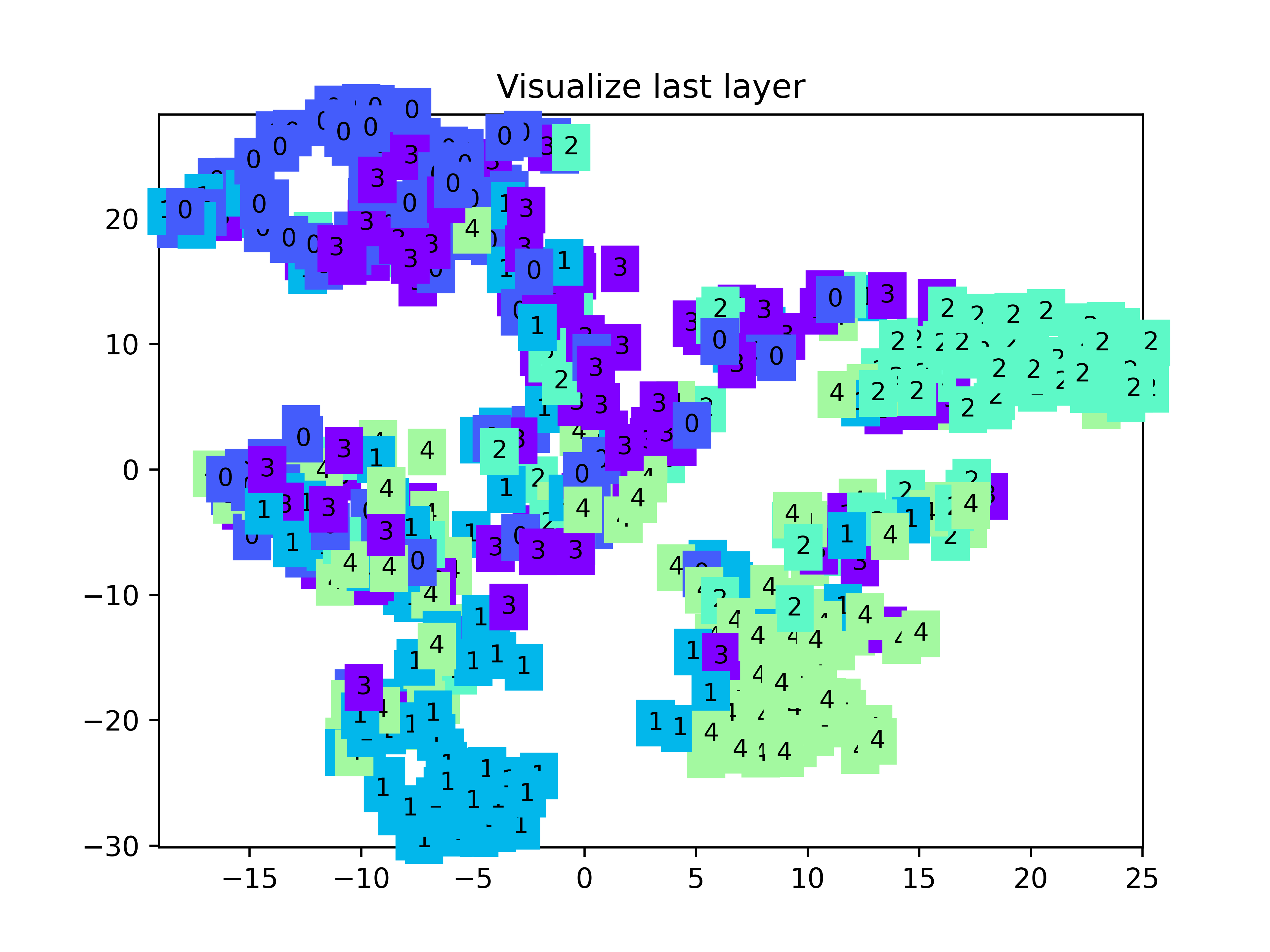}
      \caption{t-SNE visualization of the final-layer embeddings for five SARS-CoV-2 variants. Each point represents one sequence and is colored by variant label.}
      \label{fig:TSNE_CNN_Classification}
  \end{minipage}
\end{figure}

The proposed C-VAE achieved strong discriminative performance, with classification accuracy exceeding 98\% across the five SARS-CoV-2 variants on both the validation set and the independent test set. Figure~\ref{fig:Confusion_Matrix} summarizes the results using a confusion matrix. Figure~\ref{fig:TSNE_CNN_Classification} visualizes the embeddings from the final network layer using a t-SNE projection \cite{hinton2002StochasticNeighborEmbedding}.

The standardized genome sequences were processed by the encoder’s convolutional layers, producing activation maps for 12 filters. After comparing four feature extraction strategies, we selected the Pooling method due to its favorable computational efficiency while maintaining primer discovery performance. Primer appearance rates were quantified using the datasets summarized in Table~\ref{tab:SARS-CoV-2_data_selection} and Appendix Table~\ref{tab:other_Coronavirus_species}. The resulting forward primers showed high target specificity: they appeared in more than 95\% of sequences from the target variant and in less than 5\% of sequences from non-target variants. Omicron was a notable exception, with an appearance rate of approximately 80\% in target sequences and below 20\% in other variants, consistent with its higher genetic diversity and ongoing evolution \cite{Arora_2024}. For Omicron, we therefore applied a relaxed appearance-rate threshold to retain sufficiently specific primers without resorting to lineage-specific primer design.

Detailed results are reported in Table~\ref{tab:5} (the \textit{Homo sapiens} genome), Table~\ref{tab:6} (non-\textit{Homo sapiens} hosts), and Table~\ref{tab:7} (other taxa; Appendix). These validated forward primers were then used as anchors for downstream-region extraction in Stage~III, as illustrated in Appendix Figure~\ref{Flowchart_Forward}. For the complete pipeline, including \textit{in silico} PCR validation and reverse primer design, see Figure~\ref{fig:overall_pipelines} and Appendix Figure~\ref{Flowchart_Reverse}, respectively. After filtering candidates by standard primer design constraints and appearance-frequency thresholds, we obtained variant-specific forward primers: 66 (Alpha), 23 (Beta), 59 (Gamma), 52 (Delta), and 69 (Omicron).

For reverse primer design, we adapted the C-VAE pipeline to account for the downstream-sequence setting and the additional constraints of reverse primers. Specifically, we trained an independent C-VAE model for each validated forward primer and compared four feature extraction strategies (Pooling, Top-$k$, Mix, and Reconstruction). Performance was evaluated by the number of viable reverse primer candidates (and primer pairs) produced after applying the downstream filtering and screening criteria. The effectiveness of these strategies varied by variant. For the Alpha variant (Appendix Table~\ref{tab:11}), the Top-$k$, Mix, and Reconstruction methods produced substantially more viable candidates than Pooling. For the Delta variant (Appendix Table~\ref{tab:12}), the Top-$k$ method yielded the highest extraction efficiency. Based on these evaluations, we selected the Top-$k$ method as the default strategy for reverse primer extraction. We then computed the occurrence-frequency distributions of the resulting reverse primers for each variant (Appendix Table~\ref{tab:5.5}).

Final primer-pair selection was performed using thermodynamic constraints and complementarity screening. We first filtered candidates by GC content and melting temperature (Tm). We then removed primers with strong self-complementarity (self-dimers) and excluded forward--reverse combinations with substantial cross-complementarity (heterodimers). Finally, we required the Tm difference between paired primers to be within $5^{\circ}$C to support robust amplification. Applying these criteria yielded 1{,}478 computationally validated primer pairs across the five variants, summarized in Table~\ref{tab:13}.

\begin{table}[H]
\centering
\small
\begin{tabular}{c|c|c|c|c|c}
 & Forward Primer  & Reverse Primer & Amplicon Size & Amplicon Size & Amplicon Size \\
 & Number & Number & \textless 200 bp & \textless 500 bp & \textless 1,000 bp \\ \hline
Alpha & 66 & 400 & 6 & 14 & 66\\ \hline
Beta & 23 & 18 & 0 & 0 & 1 \\ \hline
Gamma & 59 & 272 & 0 & 49 & 66 \\ \hline
Delta & 52 & 457 & 33 & 106 & 154 \\ \hline
Omicron & 69 & 331 & 23 & 26 & 50 \\
\hline
\end{tabular}
\caption{Numbers of generated forward and reverse primers and computationally validated primer pairs by amplicon-length threshold for each SARS-CoV-2 variant. Reverse primers are generated conditional on each validated forward primer, and primer pairs are formed only from compatible forward--reverse combinations.}
\label{tab:13}
\end{table}

\paragraph{Primer-BLAST}

After computational design and thermodynamic screening, we performed Primer-BLAST \cite{ye2012PrimerBLASTToolDesign} as an additional specificity check within SARS-CoV-2. Specifically, Primer-BLAST searches were restricted to SARS-CoV-2 sequences only. This step was used to verify that each primer pair maps to the intended locus and does not produce unintended matches elsewhere within SARS-CoV-2 genomes. Cross-reactivity to non-SARS-CoV-2 taxa was assessed separately via the appearance-frequency analysis described above. The detailed search settings and representative alignment outputs are provided in Appendix Figures~\ref{fig:BALST} and \ref{fig:BLAST_result}, respectively.

\paragraph{\textit{in silico} PCR}

\begin{figure}[htbp]
    \centering
    \includegraphics[scale=0.6]{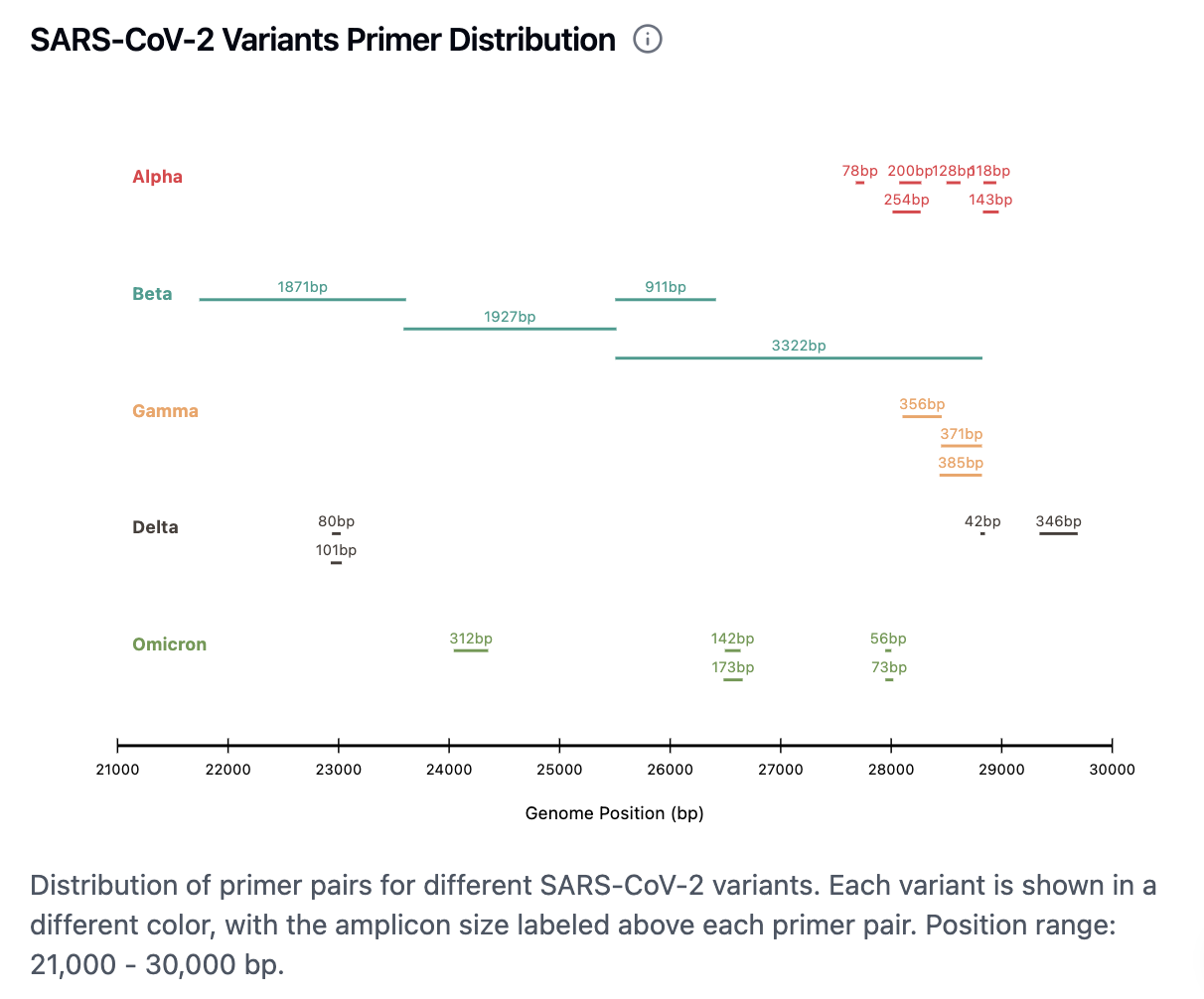}
    \caption{Visualization of primer pair binding positions in the target genome sequence.}
    \label{fig:SARS-CoV-2_Primer_Distribution}
\end{figure}

Primer pairs that passed all design and specificity criteria were then evaluated using \textit{in silico} PCR on two independent platforms: FastPCR \cite{kalendar2017FastPCRSilicoTool} and Unipro UGENE \cite{okonechnikov2012UniproUGENEUnified}. Candidate forward and reverse primers were assessed in batch mode in both tools. The results indicate that primer pairs produced by our pipeline amplify the intended variant-specific regions with high specificity. The \textit{in silico} PCR outputs include genomic binding coordinates, melting temperature (Tm), predicted complementarity, and expected amplicon length. Representative results are shown in Appendix Figures~\ref{fig:FastPCR} and \ref{fig:Unipro_UGENE}. Based on these validation steps, we report a curated set of 22 primer pairs in Table~\ref{tab:SARS-CoV-2 Primer Pairs}, with their genomic binding positions summarized in Figure~\ref{fig:SARS-CoV-2_Primer_Distribution}.

{\scriptsize{
\begin{center}
\begin{table}[H]
\centering
\begin{tabular}{c|c|c|c|c}
Primers (5' to 3') & GC content & Tm ($^{\circ}$C) &  Position & Amplicon Size\\ \hline
Alpha Variant \\ \hline
F - AGGAGCTATAAAATCAGCACC & 42.86\% & 49.60 & 27680-\textgreater27700 & \multirow{2}*{78 bps} \\
R - TCGATGCACTGAATGGGTGAT & 47.62\% & 53.89 & 27737\textless-27757 & ~  \\ \hline
F - TCAACTCCAGGCAGCAGTAAAC & 50.00\% & 54.93 & 28834-\textgreater28855 & \multirow{2}*{118 bps} \\
R - CAAACATTTTGCTCTCAAGCTG & 40.91\% & 51.34 & 28930\textless-28951 & ~  \\ \hline
F - TTCAACTCCAGGCAGCAGTAA & 47.62\% & 52.40 & 28500-\textgreater28520 & \multirow{2}*{128 bps} \\
R - GGCCTTTACCAAACATTTTGC & 42.86\% & 50.45 & 28607\textless-28627 & ~  \\ \hline
F - AATTCAACTCCAGGCAGCAGTAAAC & 44.00\% & 56.04 & 28830-\textgreater28855 & \multirow{2}*{143 bps} \\
R - CCTTGTTGTTGTTGGCCTTTACCAA & 44.00\% & 56.60 & 28948\textless-28973 & ~  \\ \hline
F - CCATTCAGTGCATCGATATCGG & 50.00\% & 53.59 & 28073-\textgreater28097 & \multirow{2}*{200 bps} \\
R - CTGATTTTGGGGTCCATTTAGA & 40.91\% & 50.11 & 28251\textless-28272 & ~  \\ \hline
F - GAGCTATAAAATCAGCACC & 42.11\% & 45.40 & 28012-\textgreater28031 & \multirow{2}*{254 bps} \\
R - TTGGGGTCCATTTAGAGACAT & 42.86\% & 50.31 & 28245\textless-28266 & ~  \\ \hline
\\ \hline
Beta Variant \\ \hline
F - TCATAGCGCTTCCAAAATC & 42.11\% & 47.81 & 25503-\textgreater25522 & \multirow{2}*{911 bps} \\
R - AGACCAGAAGATCAAGAACTCTAG & 41.67\% & 51.29 & 26390\textless-26414 & ~  \\ \hline
F - GTTTGCTAACCCTGTCCTACCAT & 47.83\% & 54.33 & 21740-\textgreater21762 & \multirow{2}*{1,871 bps} \\
R - CTACACCAAGTGACATAGTGTAG & 43.48\% & 50.36 & 23588\textless-23610 & ~  \\ \hline
F - CTACACTATGTCACTTGGTGTA & 40.91\% & 49.16 & 23588-\textgreater23609 & \multirow{2}*{1,927 bps} \\
R - AAGCGCTATGAAAAACAGCAAG & 40.91\% & 52.69 & 25493\textless-25514 & ~  \\ \hline
F - TCATAGCGCTTCCAAAATC & 42.11\% & 47.81 & 25504-\textgreater25522 & \multirow{2}*{3,322 bps} \\
R - CTACTGCTGCCTGGAGTTG & 57.89\% & 52.15 & 28807\textless-28825 & ~  \\ \hline
\\ \hline
Gamma Variant \\ \hline
F - GCCAGAAACCTAAATTGGGTA & 42.86\% & 49.96 & 28102-\textgreater28122 & \multirow{2}*{356 bps} \\
R - CATCTCGACTGCTATTGGTGT & 47.62\% & 52.05 & 28437\textless-28457 & ~  \\ \hline
F - CGAGATGACCAAATTGGCTAC & 47.62\% & 51.34 & 28451-\textgreater28471 & \multirow{2}*{371 bps} \\
R - TTAGAGCTGCCTGGAGTTGAA & 47.62\% & 53.08 & 28801\textless-28821 & ~  \\ \hline
F - ACACCAATAGCAGTCGAGATG & 47.62\% & 52.05 & 28437-\textgreater28457 & \multirow{2}*{385 bps} \\
R - TTAGAGCTGCCTGGAGTTGAA & 47.62\% & 53.08 & 28801\textless-28821 & ~  \\ \hline
\\ \hline
Delta Variant\\ \hline
F - TCAACTCCAGGCAGCAGTATG & 52.38\% & 54.20 & 28807-\textgreater28827  & \multirow{2}*{42 bps} \\
R - CATTCTAGCAGGAGAAGTTCC & 47.62\% & 50.00 & 28828\textless-28848 & ~  \\ \hline
F - GGTAGCAAACCTTGTAATGGT & 42.86\% & 50.30 & 22940-\textgreater22960  & \multirow{2}*{80 bps} \\
R - CCATTAGTGGGTTGGAAACCA & 47.62\% & 52.19 & 22999\textless-23019 & ~  \\ \hline
F - TCTATCAGGCCGGTAGCAAAC & 52.38\% & 54.02 & 22929-\textgreater22949  & \multirow{2}*{101 bps} \\
R - GTAACCAACACCATTAGTGGG & 47.62\% & 50.72 & 23009\textless-23029 & ~  \\ \hline
F - AGGCTTATGAAACTCAAGCCT & 42.86\% & 51.34 & 29342-\textgreater29362  & \multirow{2}*{346 bps} \\
R - AGTGGCCTCGGTGAAAATGTG & 52.38\% & 55.31 & 29667\textless-29687 & ~  \\ \hline
\\ \hline
Omicron Variant\\ \hline
F - CACTCCGCATTACGTTTGGTG & 52.38\% & 54.48 & 27946-\textgreater27966 & \multirow{2}*{56 bps} \\
R - ACCATTCTGGTTACTGCCAGT & 47.62\% & 53.32 & 27981\textless-28001 & ~  \\ \hline
F - ACTCCGCATTACGTTTGGTGG & 52.38\% & 55.42 & 27947-\textgreater27967 & \multirow{2}*{73 bps} \\
R - TTGTTTTGATCGCGCCCCACC & 57.14\% & 58.53 & 27999\textless-28019 & ~  \\ \hline
F - CTCCTTGAAGAATGGAACCT & 45.00\% & 48.79 & 26495-\textgreater26515 & \multirow{2}*{142 bps} \\
R - TTAAAGTTACTGGCCATAACAGCC & 41.67\% & 53.36 & 26613\textless-26637 & ~  \\ \hline
F - GAGCTTAAAAAGCTCCTTGAAG & 40.91\% & 49.90 & 26483-\textgreater26505 & \multirow{2}*{173 bps} \\
R - GCAGCAAGCACAAAACAAGTT & 42.86\% & 53.28 & 26635\textless-26656 & ~  \\ \hline
F - GTGCACAAAAGTTTAACGGCCT & 45.45\% & 52.38 & 24042-\textgreater24063 & \multirow{2}*{312 bps} \\
R - TATGGTTGACCACATCTTGAAG & 40.91\% & 50.23 & 24332\textless-24353 & ~  \\ \hline
\end{tabular}
\caption{Primer pairs successfully validated via in-silico PCR for each SARS-CoV-2 virus variant detection.}
\label{tab:SARS-CoV-2 Primer Pairs}
\end{table}
\end{center}
}
}

\subsection{Experiment 2: Primer Design for \textit{E. coli} and \textit{S. flexneri}} \label{sec:Primer design for E.coli and S. flexneri}

\textit{Escherichia coli} (\textit{E. coli}) and \textit{Shigella flexneri} (\textit{S. flexneri}) are clinically important bacterial species with distinct implications for human health. \textit{E. coli} is commonly a commensal member of the human intestinal microbiome, whereas \textit{S. flexneri} is an established pathogen associated with shigellosis and a substantial global burden of gastrointestinal disease and foodborne transmission \cite{croxen2013RecentAdvancesUnderstanding, scallan2011FoodborneIllnessAcquired}. Rapid and reliable identification of these organisms is therefore important for clinical decision-making and public health surveillance.

Although PCR-based assays are widely used to detect \textit{E. coli} and \textit{S. flexneri} due to their high analytical sensitivity and specificity \cite{wildeboer2010RapidDetectionEscherichia, chen2019RapidSensitiveDetection}, designing primers that robustly discriminate between these two species remains challenging. This difficulty is driven by substantial genomic similarity between the species, together with considerable within-species genetic diversity. To address this discrimination problem, we adapted Primer C-VAE for differential detection of \textit{E. coli} and \textit{S. flexneri} using full-length genomes.

\begin{figure}[htbp]
      \centering
      \includegraphics[scale=0.6]{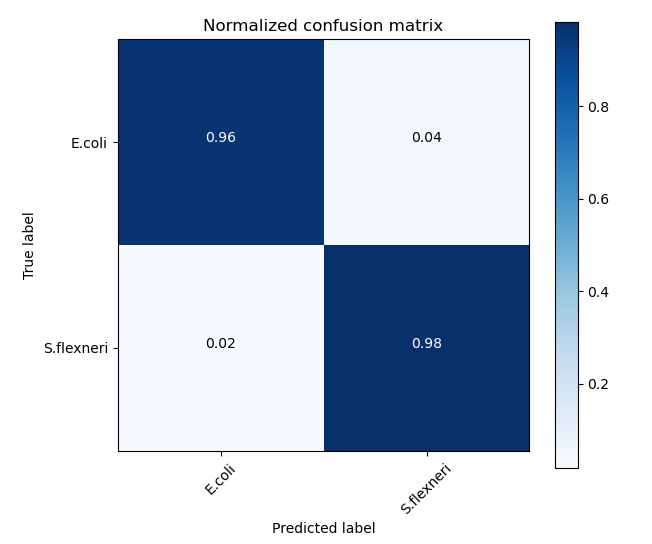}
      \caption{Confusion matrix from cross-validation for the C-VAE model based on 1,000 sequences of each E. coli and S. flexneri.}
      \label{fig:E.coli_and_S.flexneri_confusion_matrix}
\end{figure}

Our analysis included 8{,}939 complete genome sequences for \textit{E. coli} and 5{,}373 for \textit{S. flexneri}. For model development and evaluation, we constructed balanced datasets from NCBI by sampling 1{,}000 sequences per species for training, 1{,}000 per species for validation, and 1{,}000 per species for testing (Table~\ref{tab:E.coli_and_S.flexneri_dataset}). Using this setup, the optimised C-VAE achieved classification accuracy exceeding 97\% on both the validation set and the independent test set. Detailed performance metrics are reported in Table~\ref{tab:E.coli_and_S.flexneri_dataset}, and the corresponding confusion matrix is shown in Figure~\ref{fig:E.coli_and_S.flexneri_confusion_matrix}.

\begin{table}[H]
  \centering
  \begin{tabular}{c|ccc|cc}
   & Training set & Validation set & Test set & Generated primers & Calculated appearance\\ \hline
  Source & NCBI & NCBI & NCBI & NCBI &  NCBI \\ \hline
  \textit{E. coli} & 1,000 & 1,000 & 1,000 & 800 (or) & 1,500 \\
  \textit{S. flexneri} & 1,000 & 1,000 & 1,000 & 800 (or) & 1,500 \\ \hline
  Total Number & 2,000 & 2,000 & 2,000 & 1,600 & 3,000 \\ \hline
  \end{tabular}
  \caption{Data selection for the C-VAE model and datasets used to generate and compute appearance rates of forward primers for \textit{E. coli} and \textit{S. flexneri}.}
  \label{tab:E.coli_and_S.flexneri_dataset}
\end{table}

\begin{figure}[htbp]
  \centering
  \includegraphics[scale=0.6]{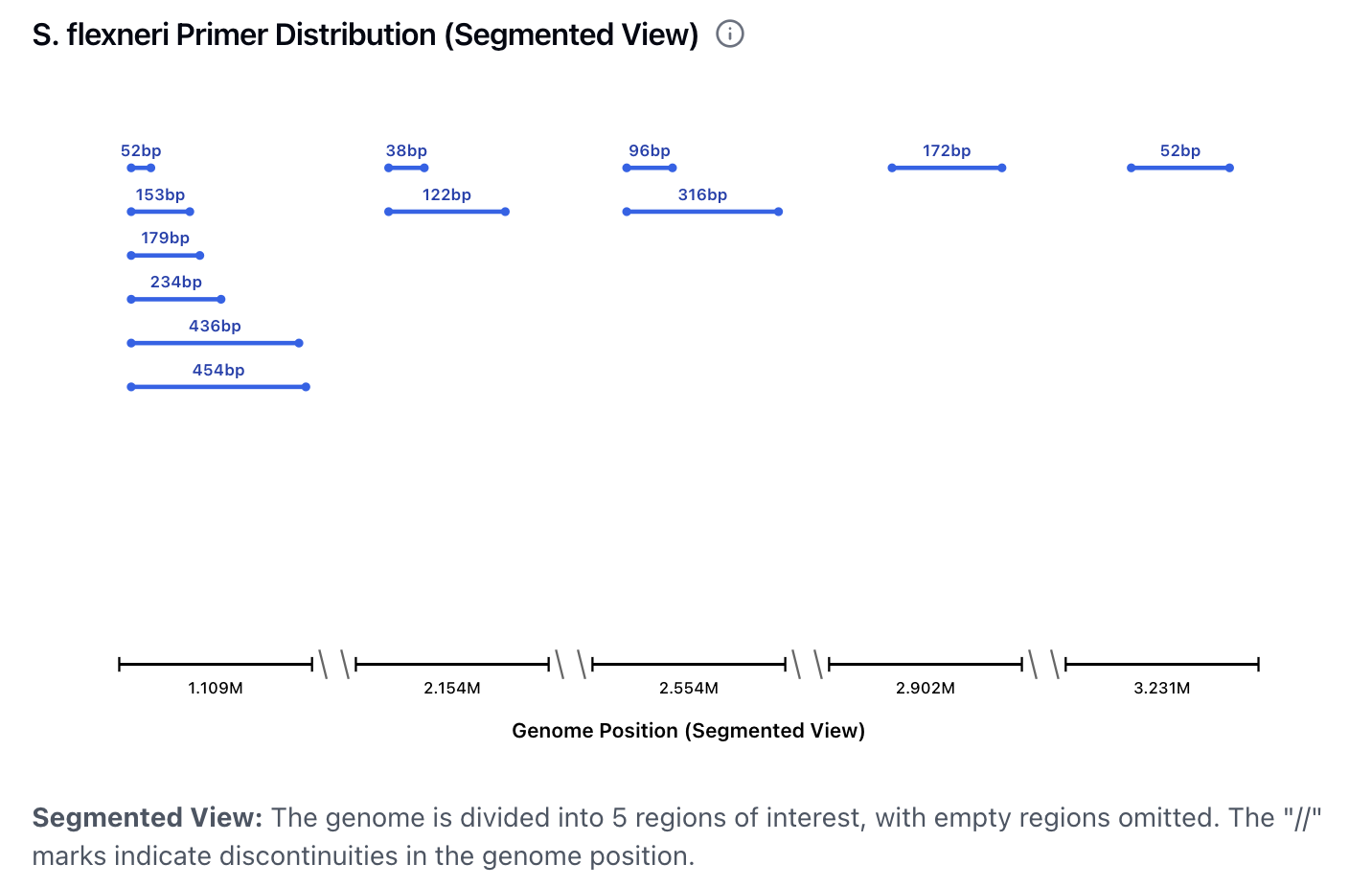}
  \caption{Visualization of primer pair binding positions in the target genome sequence.}
  \label{fig:S.flexneri_Primer_Distribution}
\end{figure}

However, because most \textit{E. coli} strains are non-pathogenic commensals and only a subset of lineages are associated with virulence, we prioritized primer-pair reporting for \textit{S. flexneri} in this experiment. The genomes of \textit{E. coli} and \textit{S. flexneri} are approximately 4.5--5.5~Mb, which is substantially larger than the SARS-CoV-2 genome (\textasciitilde30~kb). This increase in sequence length and genomic complexity poses additional computational demands for our C-VAE implementation, requiring greater memory and longer processing time than the SARS-CoV-2 experiments. Despite these challenges, the proposed pipeline produced highly specific primer candidates. The optimised \textit{S. flexneri} primer pairs are reported in Table~\ref{tab:S.flexneri Primer Pairs}, and their genomic binding coordinates are summarized in Figure~\ref{fig:S.flexneri_Primer_Distribution}.

{\scriptsize{
\begin{center}
\begin{table}[H]
\centering
\begin{tabular}{c|c|c|c|c}
Primers (5' to 3') & GC content & Tm ($^{\circ}$C) & Position & Amplicon Size\\ \hline
F -GAGCTGATGGCTTCATCCAGA & 52.38\% & 57.74 & 2153834-\textgreater2153854 & \multirow{2}*{38 bps} \\
R - GCCGATCCCCTGAAAGC & 64.71\% & 55.63 & 2153855\textless-2153871 & ~ \\ \hline

F - GTGACGCTGTAGATGATACGT & 47.62\% & 55.53 & 3230534-\textgreater3230554 & \multirow{2}*{52 bps} \\
R - AAAACCGTCTGAAAAGCCGCA & 47.62\% & 59.49 & 3230565\textless-3230585 & ~ \\ \hline

F - GCTTTAGTATCGACTTGCTGA & 42.86\% & 53.67 & 1109031-\textgreater1109051 & \multirow{2}*{52 bps} \\
R - TATTGCTGGGTAATCAGGCGT & 47.62\% & 57.38 & 1109062\textless-1109082 & ~ \\ \hline

F - GCGCGGTTTTAATGAAGAAGA & 42.86\% & 55.39 & 2554171-\textgreater2554191 & \multirow{2}*{96 bps} \\
R - TCACGACGCATCAGATGATGC & 52.38\% & 59.02 & 2554246\textless-2554266 & ~ \\ \hline

F - GAGCTGATGGCTTCATCCAGA & 52.38\% & 57.75 & 2153834-\textgreater2153854 & \multirow{2}*{122 bps} \\
R - GTGACTAACGGCAGCGGTAAG & 57.14\% & 59.23 & 2153935\textless-2153955 & ~ \\ \hline

F - GCTTTAGTATCGACTTGCTGA & 42.86\% & 53.67 & 1109031-\textgreater1109051 & \multirow{2}*{153 bps} \\
R - AGCGATCCTTGATAAAGCAGG & 47.62\% & 56.02 & 1109163\textless-1109183 & ~ \\ \hline

F - AGCTCAAGCAACAATTTACGC & 42.86\% & 55.92 & 2902098-\textgreater2902118 & \multirow{2}*{172 bps} \\
R - TTAAAGCTCTTGCCGCAGAGG & 52.38\% & 58.83 & 2902249\textless-2902269 & ~ \\ \hline

F - GCTTTAGTATCGACTTGCTGA & 42.86\% & 53.67 & 1109031-\textgreater1109051 & \multirow{2}*{179 bps} \\
R - CAAGGTTCCAGCCATCCATTG & 52.38\% & 57.41 & 1109190\textless-1109209 & ~ \\ \hline

F - GCTTTAGTATCGACTTGCTGA & 42.86\% & 53.67 & 1109031-\textgreater1109051 & \multirow{2}*{234 bps} \\
R - AACATTTGCTGATGTTGACGA & 38.1\% & 54.57 & 1109244\textless-1109264 & ~ \\ \hline

F - GCGCGGTTTTAATGAAGAAGA & 42.86\% & 55.39 & 2554171-\textgreater2554191 & \multirow{2}*{316 bps} \\
R - TTGTGCCTGTAATGTGGTGCC & 52.38\% & 59.31 & 2554466\textless-2554486 & ~ \\ \hline

F - GCTTTAGTATCGACTTGCTGA & 42.86\% & 53.67 & 1109031-\textgreater1109051 & \multirow{2}*{436 bps} \\
R - CTACGGTGCTGATTATCGCCT & 52.38\% & 57.89 & 1109446\textless-1109466 & ~ \\ \hline

F - GCTTTAGTATCGACTTGCTGA & 42.86\% & 53.67 & 1109031-\textgreater1109051 & \multirow{2}*{454 bps} \\
R - ATACCCTGGTGATTGCCACTA & 47.62\% & 56.38 & 1109464\textless-1109484 & ~ \\ \hline
\end{tabular}
\caption{Primer pairs selected from S. flexneri with amplicon sizes ranging from 0-500 bps, sorted by amplicon size.}
\label{tab:S.flexneri Primer Pairs}
\end{table}
\end{center}
}}

\section{Discussion} \label{section_Final_discussion}

We propose a VAE-based deep-learning workflow for flexible-length primer design that supports target-specific detection across diverse organisms. The approach addresses a common practical challenge in primer development: identifying highly discriminative primer regions in settings with substantial genetic heterogeneity, such as SARS-CoV-2. In current practice, variant-oriented primer design for SARS-CoV-2 often relies on manual screening of large collections of full-length genomes to identify variant-defining alterations (e.g., deletions \cite{vogels2021MultiplexQPCRDiscriminates} or characteristic mutations \cite{sibai2022DevelopmentEvaluationRTqPCR, wang2021MultiplexSARSCoV2Genotyping}). Such workflows are time-consuming and typically require substantial domain expertise. In addition, widely used automated tools such as Primer3 and Primer3Plus \cite{koressaar2007EnhancementsModificationsPrimer, untergasser2012Primer3newCapabilitiesInterfaces} impose practical constraints for long-genome applications (e.g., a maximum input length below 10{,}000~bp; Appendix Table~\ref{tab: Primer3_vs_C-VAE}). These constraints limit direct application to organisms with large genomes, including \textit{E. coli} and \textit{S. flexneri} (4.5--5.5~Mb), and may require truncation or preprocessing that can discard potentially informative regions. By contrast, Primer C-VAE is designed to reduce manual screening effort and support long sequences. As shown in Table~\ref{tab: Primer3_vs_C-VAE} (Delta example: hCoV-19/Indonesia/JK-GS-FKUINIHRD-0489/2022), our workflow improves the efficiency of identifying candidate primer regions from complete genomes.

Across our experiments, \textit{in silico} PCR validation indicates that primer pairs produced by the proposed pipeline achieve high specificity for the intended targets. The C-VAE encoder learns discriminative sequence representations that support sequence classification and guide primer candidate extraction, while the decoder provides an auxiliary reconstruction objective that regularizes representation learning. Differences between the input and reconstructed sequences can also help highlight regions that vary systematically across classes, offering a degree of interpretability that is useful for downstream primer design. Together, these components demonstrate a practical integration of deep learning and primer design workflows for variant- and species-level discrimination.

Several limitations and opportunities for improvement remain. First, the current pipeline does not incorporate degenerate bases, which could improve robustness to within-class variability and represents an important extension for future work. Second, although we apply thermodynamic and complementarity screening (e.g., GC content, melting temperature, and dimer checks) prior to \textit{in silico} PCR, these steps still require multiple filters and tool-based evaluations; further automation and tighter integration of these criteria could streamline the workflow. Third, the computational cost of training remains non-trivial, particularly for reverse primer design where separate models are trained conditional on each validated forward primer and additional hyperparameter tuning may be required. Fourth, a limitation of this study is the absence of wet-lab experimental validation. Although our \textit{in silico} PCR validation using multiple established tools (FastPCR, Unipro UGENE, and Primer-BLAST) provides strong computational evidence for primer specificity, experimental confirmation of amplification performance remains an important step before clinical or field deployment; future work will focus on laboratory validation of selected primer pairs. Finally, the method depends on having sufficient genome sequences for training; for organisms with limited publicly available complete genomes (e.g., Human astrovirus, HAstV), direct application may be constrained. Addressing these challenges---for example through more efficient training strategies, stronger sharing of information across primer-specific models, and data-efficient learning---is a promising direction for future research.

\section*{Acknowledgement}

The work of Hanyu Wang is funded by China Scholarship Council with Reference 202308060184. The work of Anthony J. Dunn is jointly funded by Decision Analysis Services Ltd and EPSRC through the studentship with Reference EP/R513325/1. The work of Alain B. Zemkoho is supported by the EPSRC grant EP/V049038/1 and The Alan Turing Institute under the EPSRC grants EP/N510129/1 and EP/W037211/1.

\section*{Conflict of interest statement}

The authors declare that there is no conflict of interest.

\section{Hardware and Software Environments}

All experiments were implemented in Python and evaluated in the following hardware and software environments.

\subsection*{Windows/Linux (GPU-accelerated)}
\begin{itemize}
    \item Operating system: Windows 11 Pro 23H2 and Ubuntu 22.04.5 LTS
    \item CPU: 13th Gen Intel(R) Core(TM) i7-13700 @ 2.10\,GHz
    \item RAM: 32.0\,GB
    \item GPU: NVIDIA GeForce RTX 4070 Ti
    \item Python: 3.9.7 (64-bit)
    \item PyTorch: 2.5.0 with CUDA 12.4
    \item IDE: PyCharm 2024.2.3 (Professional Edition)
\end{itemize}

\subsection*{macOS (CPU-only)}
We additionally tested the pipeline on macOS without GPU acceleration (CPU-only execution).
\begin{itemize}
    \item Operating system: macOS Sonoma 14.4
    \item CPU: Apple M1 (8 cores: 4 performance + 4 efficiency)
    \item GPU: Apple M1 7-core GPU
    \item RAM: 8.0\,GB
    \item Python: 3.12.4
    \item PyTorch: 2.5.0
    \item IDE: PyCharm 2024.2.3 (Professional Edition)
\end{itemize}

\section*{Data availability}
We gratefully acknowledge the authors responsible for obtaining the specimens and genetic sequence data generated and shared via the GISAID Initiative (\href{https://doi.org/10.55876/gis8.220628xf}{https://doi.org/10.55876/gis8.220628xf}), and that we used for the research presented in this paper. All the codes used and generated in the course of the work presented in this article are available in the following GitHub page: \href{https://github.com/harrywang9917/Primer_C-VAE}{https://github.com/harrywang9917/Primer\_C-VAE}


\bibliographystyle{apalike}
\normalem


\newpage

\begin{appendices}

\section{Data collection and pre-processing} \label{Appendix_1}

This appendix provides supplementary details on data collection and preprocessing. Table~\ref{tab:SARS-COV-2_samples} summarizes the SARS-CoV-2 variant metadata used in this study, including WHO labels, Pango lineages, GISAID clades, sample counts, and the corresponding class labels used in our experiments. Table~\ref{tab:SARS-CoV-2_seq_length} reports the average genome length for each SARS-CoV-2 variant. Figure~\ref{fig:sequence_encoding_example} illustrates an example of sequence standardization and encoding. Table~\ref{tab:other_Coronavirus_species} lists the SARS-CoV-2 and non-SARS-CoV-2 coronavirus sequences collected from GISAID and NCBI that were used to compute the appearance rates of forward and reverse primers.

\begin{table}[H]
\centering
\begin{tabular}{ccccc}
\textbf{WHO label} & \textbf{Pango lineage} & \textbf{GISAID clade} & \textbf{Number of samples} & \textbf{Label} \\ \hline
Alpha   & B.1.1.7+Q.*       & GRY         & 119{,}077 & 1 \\
Beta    & B.1.351           & GH/501Y.V2  & 27{,}782  & 2 \\
Gamma   & P.1               & GR/501Y.V3  & 48{,}588  & 3 \\
Delta   & B.1.617.2+AY.*    & G/478K.V1   & 142{,}815 & 4 \\
Omicron & B.1.1.529+BA.*    & GR/484A     & 135{,}383 & 0 \\ \hline
\end{tabular}
\caption{SARS-CoV-2 variants collected from GISAID for C-VAE model development.}
\label{tab:SARS-COV-2_samples}

{\raggedright \textbf{Note:} SARS-CoV-2 variants are described under multiple nomenclature systems. This table reports the WHO label, Pango lineage, and GISAID clade for each variant, together with the sample counts and the class labels used in this study. \par}
\end{table}

\begin{table}[H]
\centering
\begin{tabular}{ccc}
\textbf{Variant} & \textbf{Average length (bp)} & \textbf{Label} \\ \hline
Alpha   & 29{,}769 & 1 \\
Beta    & 29{,}774 & 2 \\
Gamma   & 29{,}770 & 3 \\
Delta   & 29{,}766 & 4 \\
Omicron & 29{,}748 & 0 \\ \hline
\end{tabular}
\caption{Average genome sequence length for each SARS-CoV-2 variant.}
\label{tab:SARS-CoV-2_seq_length}

{\raggedright \textbf{Note:} The average genome lengths of the five variants are very similar. In our dataset, the longest sequence was a Delta genome with length 31{,}079~bp. \par}
\end{table}

\begin{figure}[htbp]
    \centering
    \begin{minipage}{0.95\textwidth}
        \textbf{Example of sequence standardization and ordinal encoding:}\\[2mm]
        1) Original sequences with different lengths:\\
        {\color{sequenceRed}
        5'- AGTCAGCATCTCATGTGCGAGTCCTGACGCTGACTAGC -3'} (38~bp)\\
        {\color{sequenceRed}
        5'- ATCTCATGTGCGAACGCTGACTAGAAAATCCAAAAAANNNNNNA -3'} (45~bp)\\
        {\color{sequenceRed}
        5'- CGCTGACTAGAAAATCCAAAAAANNNCGTTTACTTCGANNN -3'} (41~bp)\\
        {\color{sequenceRed}
        5'- NNNNAGTCAGCATCTCATGTGTCCTGACGCTGACTAG -3'} (37~bp)\\[2mm]

        2) Standardized sequences (padded with N to the maximum length):\\
        {\color{sequenceBlue}
        5'- AGTCAGCATCTCATGTGCGAGTCCTGACGCTGACTAGC}{\color{sequenceYellow}NNNNNNN} -3' (45~bp)\\
        {\color{sequenceBlue}
        5'- ATCTCATGTGCGAACGCTGACTAGAAAATCCAAAAAANNNNNNA -3'} (45~bp)\\
        {\color{sequenceBlue}
        5'- CGCTGACTAGAAAATCCAAAAAANNNCGTTTACTTCGANNN}{\color{sequenceYellow}NNNN} -3' (45~bp)\\
        {\color{sequenceBlue}
        5'- NNNNAGTCAGCATCTCATGTGTCCTGACGCTGACTAG}{\color{sequenceYellow}NNNNNNNN} -3' (45~bp)\\[2mm]

        3) Ordinal encoding:\\
        \begin{tabular}{@{}l@{ }l@{}}
        Encoded: & 4 3 2 1 4 3 1 4 2 1 2 1 4 2 3 2 3 1 3 4 3 2 1 1 2 3 4 1 3 1 2 3 4 1 2 4 3 1 0 0 0 0 0 0 0\\
        Encoded: & 4 2 1 2 1 4 2 3 2 3 1 3 4 4 1 3 1 2 3 4 1 2 4 3 4 4 4 4 2 1 1 4 4 4 4 4 4 0 0 0 0 0 0 0 4\\
        Encoded: & 1 3 1 2 3 4 1 2 4 3 4 4 4 4 2 1 1 4 4 4 4 4 4 0 0 0 1 3 2 2 2 4 1 2 2 1 3 4 0 0 0 0 0 0 0\\
        Encoded: & 0 0 0 0 4 3 2 1 4 3 1 4 2 1 2 1 4 2 3 2 3 2 1 1 2 3 4 1 3 1 2 3 4 1 2 4 3 1 0 0 0 0 0 0 0
        \end{tabular}
    \end{minipage}
    \caption{Example of sequence standardization and ordinal encoding. Sequences are padded with ``N'' to a fixed length and then mapped to integers using A=4, G=3, T=2, C=1, and N=0.}
    \label{fig:sequence_encoding_example}
\end{figure}

\begin{table}[H]
\centering
\begin{tabular}{cccc}
\textbf{Coronavirus species} & \textbf{Source} & \textbf{Host} & \textbf{Number of samples} \\ \hline
SARS-CoV-2 (all five variants) & GISAID & \textit{Homo sapiens} & 58{,}547 \\
SARS-CoV-2 (all five variants) & NCBI   & \textit{Homo sapiens} & 78{,}692 \\
SARS-CoV-2                    & GISAID & \textit{Manis javanica} & 19 \\
SARS-CoV-2                    & GISAID & \textit{Rhinolophus affinis} & 1 \\
SARS-CoV-2                    & GISAID & \textit{Rhinolophus} & 1 \\
SARS-CoV-2                    & GISAID & \textit{Canis} & 29 \\
SARS-CoV-2                    & GISAID & \textit{Felis catus} & 51 \\
MERS-CoV                      & NCBI   & \textit{Homo sapiens} & 738 \\
HCoV-OC43                     & NCBI   & \textit{Homo sapiens} & 1{,}311 \\
HCoV-NL63                     & NCBI   & \textit{Homo sapiens} & 634 \\
HCoV-229E                     & NCBI   & \textit{Homo sapiens} & 446 \\
HCoV-HKU1                     & NCBI   & \textit{Homo sapiens} & 404 \\
SARS-CoV-P2                   & NCBI   & \textit{Homo sapiens} & 1 \\
SARS-CoV-HKU-39849            & NCBI   & \textit{Homo sapiens} & 2 \\
SARS-CoV-GDH-BJH01            & NCBI   & \textit{Homo sapiens} & 1 \\
HAstV-BF34                    & NCBI   & \textit{Homo sapiens} & 2 \\ \hline
\end{tabular}
\caption{Coronavirus sequences used to compute primer appearance rates.}
\label{tab:other_Coronavirus_species}

{\raggedright \textbf{Note:} In addition to SARS-CoV-2 variant genomes, we included SARS-CoV-2 sequences from non-human hosts and other coronavirus species to evaluate cross-reactivity when computing the appearance rates of generated primers. \par}
\end{table}

\section{Feature extraction and evaluation} \label{Appendix_2}

This appendix provides supplementary results for feature extraction and primer evaluation. Table~\ref{tab:Top_method} reports the effect of different Top-$k$ values on the number of primer pairs obtained and the resulting amplicon-length ranges. Tables~\ref{tab:5} and \ref{tab:6} summarize forward primer appearance rates evaluated on \textit{Homo sapiens} and non-\textit{Homo sapiens} hosts, respectively. Table~\ref{tab:7} reports forward primer appearance rates across other coronavirus taxa. Tables~\ref{tab:11} and \ref{tab:12} present the numbers of reverse primer candidates generated per forward primer under the four extraction methods (Pooling, Top-$k$, Mix, and Reconstruction) for the Alpha and Delta variants, as described in Section~\ref{sec:Reverse_primer_design}. Finally, Table~\ref{tab:5.5} reports reverse primer appearance rates evaluated on \textit{Homo sapiens} hosts, analogous to the forward-primer analysis in Table~\ref{tab:5}.

\begin{table}[H]
\centering
\begin{tabular}{c|cccc}
\textbf{Variant / metric} & \textbf{Top-75} & \textbf{Top-125} & \textbf{Top-175} & \textbf{Top-250} \\ \hline
\textbf{Alpha: total primer pairs} & 60 & 94 & 136 & 171 \\ \hline
Amplicon length $<1{,}000$~bp & 7 & 11 & 17 & 17 \\
Amplicon length $<500$~bp     & 7 & 10 & 16 & 15 \\
Amplicon length $<300$~bp     & 0 & 0 & 0 & 0 \\
Amplicon length $<200$~bp     & 0 & 0 & 0 & 0 \\ \hline
\textbf{Delta: total primer pairs} & 290 & 483 & 623 & 736 \\ \hline
Amplicon length $<1{,}000$~bp & 0 & 1 & 4 & 2 \\
Amplicon length $<500$~bp     & 0 & 0 & 0 & 0 \\
Amplicon length $<300$~bp     & 0 & 0 & 0 & 0 \\
Amplicon length $<200$~bp     & 0 & 0 & 0 & 0 \\ \hline
\end{tabular}
\caption{Effect of the Top-$k$ parameter on the number of generated primer pairs and their amplicon-length distribution for the Alpha and Delta variants.}
\label{tab:Top_method}

{\raggedright \textbf{Note:} The performance of the Top-$k$ feature extraction method depends on the choice of $k$. \par}
\end{table}

\begin{table}[H]
\centering
\small
\begin{tabular}{c|ccccc}
\multirow{2}*{\textbf{Forward primer (5$'$--3$'$)}} & \textbf{SARS-CoV-2} & \textbf{SARS-CoV-2} & \textbf{SARS-CoV-2} & \textbf{SARS-CoV-2} & \textbf{SARS-CoV-2}\\
~ &(Alpha) &(Beta) &(Gamma) & (Delta) &(Omicron)\\
\hline
\multirow{2}*{Dataset} & GISAID & GISAID  & GISAID  & GISAID  & GISAID \\
~ & and NCBI & and NCBI & and NCBI & and NCBI & and NCBI\\
Host & \textit{Homo sapiens} & \textit{Homo sapiens} & \textit{Homo sapiens} & \textit{Homo sapiens} & \textit{Homo sapiens}\\
Number of sequences & 5{,}000 & 5{,}000 & 5{,}000 & 5{,}000 & 5{,}000\\ \hline
Alpha Variant & & & & & \\ \hline
TACTAATGATAACACCTCAAG & 0.9928 & 0.0038 & 0.0004 & 0.0 & 0.0 \\
CAATTTGGCAGAGACATTGAT & 0.9928 & 0.0004 & 0.0004 & 0.0 & 0.0 \\
TCAAACTGTCAAACCTGGTAA & 0.9926 & 0.001 & 0.0008 & 0.0002 & 0.0002 \\
CTTTTCAAACTGTCAAACCTG & 0.9924 & 0.001 & 0.0008 & 0.0002 & 0.0002 \\ \hline
Beta Variant & & & & & \\ \hline
CGAACAAACTAAAATGTCTGA & 0.0014 & 0.9832 & 0.0482 & 0.0304 & 0.0016 \\
GCTTAGGGTTGATACAGCCAA & 0.0142 & 0.9756 & 0.0134 & 0.0068 & 0.0056 \\
TAGGGTTGATACAGCCAATCC & 0.0142 & 0.975 & 0.0132 & 0.0068 & 0.0056 \\
AGGGTTGATACAGCCAATCCT & 0.0142 & 0.975 & 0.0132 & 0.0068 & 0.0054 \\ \hline
Gamma Variant & & & & & \\ \hline
TGTGGTAAACAAGCTACACAA & 0.0 & 0.0004 & 0.9958 & 0.0 & 0.0 \\
GTGGTAAACAAGCTACACAAT & 0.0 & 0.0004 & 0.9958 & 0.0 & 0.0 \\
GTGTGGTAAACAAGCTACACA & 0.0 & 0.0004 & 0.9954 & 0.0 & 0.0 \\
ACACAATATCTAGTACAACAG & 0.0002 & 0.0004 & 0.9934 & 0.0 & 0.0 \\ \hline
Delta Variant & & & & & \\ \hline
GATACTAGTTTGTCTGGTTTT & 0.0016 & 0.0382 & 0.0032 & 0.998 & 0.1762 \\
AGTTTGTCTGGTTTTAAGCTA & 0.0016 & 0.0382 & 0.0032 & 0.998 & 0.1766 \\
TATGGTTGATACTAGTTTGTC & 0.0046 & 0.0426 & 0.0082 & 0.9974 & 0.1764 \\
TGGTTGATACTAGTTTGTCTG & 0.0024 & 0.0408 & 0.0056 & 0.9972 & 0.1766 \\ \hline
Omicron Variant & & & & & \\ \hline
GCGCTTCCAAAATCATAACTC & 0.0004 & 0.0002 & 0.0002 & 0.0004 & 0.8344 \\
TCACACCGGAAGCCAATATGG & 0.0002 & 0.0 & 0.0 & 0.0002 & 0.833 \\
AATAACAGTCACACCGGAAGC & 0.0002 & 0.0 & 0.0 & 0.0002 & 0.8328 \\
AGAGATAGGTACGTTAATAGT & 0.0034 & 0.0004 & 0.0006 & 0.0002 & 0.8318 \\
\hline
\end{tabular}
\caption{Appearance frequencies of selected forward primers across SARS-CoV-2 variants, evaluated on 5{,}000 \textit{Homo sapiens} host sequences per variant (combined GISAID and NCBI datasets).}
\label{tab:5}

 {\raggedright \textbf{Note:} Values report the fraction of sequences in which each forward primer occurs within the corresponding variant-specific dataset. Primers are designed to be variant-discriminative; consequently, they show high occurrence in the target variant and low occurrence in non-target variants. \par}

\end{table}

\begin{sidewaystable}[]
\centering
\begin{tabular}{c|ccccc}
Forward primer (5$'$--3$'$) & SARS-CoV-2 & SARS-CoV-2 & SARS-CoV-2 & SARS-CoV-2 & SARS-CoV-2 \\ \hline
Dataset & GISAID & GISAID & GISAID & GISAID & GISAID \\
Host & \textit{Manis javanica} & \textit{Rhinolophus affinis} & \textit{Rhinolophus} & \textit{Canis} & \textit{Felis catus} \\
Number of sequences & 19 & 1 & 1 & 29 & 51 \\ \hline
 \\
 CTCAGACTAATTCTCGTCGGC & 0.0 & 0.0 & 0.0 & 0.0 & 0.0 \\
 \\
 ACTAATTCTCGTCGGCGGGCA & 0.0 & 0.0 & 0.0 & 0.0 & 0.0 \\
 \\
 TCAGACTAATTCTCGTCGGCG & 0.0 & 0.0 & 0.0 & 0.0 & 0.0 \\
 \\
 AGACTAATTCTCGTCGGCGGG & 0.0 & 0.0 & 0.0 & 0.0 & 0.0 \\
 \\
 CAGACTAATTCTCGTCGGCGG & 0.0 & 0.0 & 0.0 & 0.0 & 0.0 \\
 \\
 AACTCCAGGCAGCAGTATGGG & 0.0 & 0.0 & 0.0 & 0.0 & 0.0 \\
 \\
 ACTCCAGGCAGCAGTATGGGA & 0.0 & 0.0 & 0.0 & 0.0 & 0.0 \\
 \\
 CTCCAGGCAGCAGTATGGGAA & 0.0 & 0.0 & 0.0 & 0.0 & 0.0 \\
 \\
 CCAGGCAGCAGTATGGGAACT & 0.0 & 0.0 & 0.0 & 0.0 & 0.0 \\
 \\
 AGGCAGCAGTATGGGAACTTC & 0.0 & 0.0 & 0.0 & 0.0 & 0.0 \\
 \\
\hline
\end{tabular}
\caption{Appearance frequencies of selected forward primers in SARS-CoV-2 sequences from non-\textit{Homo sapiens} hosts (GISAID).}
\label{tab:6}

 {\raggedright \textbf{Note:} This table reports appearance frequencies for a subset of Delta forward primers evaluated on SARS-CoV-2 genomes from non-human hosts (as listed in the header). The primers were designed using human-host Delta sequences, and no matches were observed in these non-human host datasets (all reported frequencies are 0.0). \par}

\end{sidewaystable}

\begin{sidewaystable}[]
\centering
\begin{tabular}{c|ccccccccc}
\multirow{2}*{Forward primer (5$'$--3$'$)} & \multirow{2}*{MERS-CoV} & \multirow{2}*{HCoV-OC43} & \multirow{2}*{HCoV-NL63} & HCoV & HCoV & SARS-CoV & SARS-CoV & SARS-CoV & HAstV\\
~ & ~ & ~ & ~ & -229E & -HKU1 & -P2 & -HKU-39849 & -GDH-BJH01 & -BF34 \\ \hline
Dataset & NCBI & NCBI & NCBI & NCBI & NCBI & NCBI & NCBI & NCBI & NCBI\\
Host: HS (\textit{Homo sapiens}) & HS & HS & HS & HS & HS & HS  & HS & HS & HS \\
Sequence Number & 738 & 1,311 & 634 & 446 & 404 & 1 & 2 & 1 & 2 \\ \hline
\\
CTCAGACTAATTCTCGTCGGC & 0.0 & 0.0 & 0.0 & 0.0 & 0.0 & 0.0 & 0.0 & 0.0 & 0.0\\
\\
ACTAATTCTCGTCGGCGGGCA & 0.0 & 0.0 & 0.0 & 0.0 & 0.0 & 0.0 & 0.0 & 0.0 & 0.0\\
\\
TCAGACTAATTCTCGTCGGCG & 0.0 & 0.0 & 0.0 & 0.0 & 0.0 & 0.0 & 0.0 & 0.0 & 0.0\\
\\
AGACTAATTCTCGTCGGCGGG & 0.0 & 0.0 & 0.0 & 0.0 & 0.0 & 0.0 & 0.0 & 0.0 & 0.0\\
\\
CAGACTAATTCTCGTCGGCGG & 0.0 & 0.0 & 0.0 & 0.0 & 0.0 & 0.0 & 0.0 & 0.0 & 0.0\\
\\
AACTCCAGGCAGCAGTATGGG & 0.0 & 0.0 & 0.0 & 0.0 & 0.0 & 0.0 & 0.0 & 0.0 & 0.0\\
\\
ACTCCAGGCAGCAGTATGGGA & 0.0 & 0.0 & 0.0 & 0.0 & 0.0 & 0.0 & 0.0 & 0.0 & 0.0\\
\\
CTCCAGGCAGCAGTATGGGAA & 0.0 & 0.0 & 0.0 & 0.0 & 0.0 & 0.0 & 0.0 & 0.0 & 0.0\\
\\
CCAGGCAGCAGTATGGGAACT & 0.0 & 0.0 & 0.0 & 0.0 & 0.0 & 0.0 & 0.0 & 0.0 & 0.0\\
\\
AGGCAGCAGTATGGGAACTTC & 0.0 & 0.0 & 0.0 & 0.0 & 0.0 & 0.0 & 0.0 & 0.0 & 0.0\\
\\
\hline
\end{tabular}
\caption{Appearance frequencies of selected Delta forward primers evaluated on \textit{Homo sapiens} coronavirus genomes from non-SARS-CoV-2 taxa (NCBI).}
\label{tab:7}

{\raggedright \textbf{Note:} This table reports appearance frequencies for a subset of forward primers designed for the SARS-CoV-2 Delta variant, evaluated against \textit{Homo sapiens} sequences from other coronavirus species. No matches were observed in these non-SARS-CoV-2 datasets (all reported frequencies are 0.0), supporting the specificity of the primers with respect to the examined taxa. \par}

\end{sidewaystable}

\begin{table}[H]
\centering
\small
\begin{tabular}{c|c|c|c|c}
Forward Primer (5$'$--3$'$) & Alpha & Alpha & Alpha & Alpha \\
 & Pooling Method & Top method & Mix Method & Recon Method \\ \hline
TTGGCAGAGACATTGATGACA & 30 & 96 & 58 & 63\\ \hline
TCTTATGGGTTGGGATTATCC & 48 & 61	& 115 & 67\\ \hline
TTGCACGTCTTGACAAAGTTG & 37 & 60	& 52 & 43\\ \hline
TCCTTGCACGTCTTGACAAAG & 40 & 51	& 43 & 46\\ \hline
TGCACGTCTTGACAAAGTTGA & 34 & 63	& 38 & 33\\ \hline
TTTGGCAGAGACATTGATGAC & 47 & 84	& 56 & 56\\ \hline
CACAACACATTTGTGTCTGGT & 26 & 53	& 38 & 28\\ \hline
CTTGCACGTCTTGACAAAGTT & 42 & 57	& 49 & 50\\ \hline
CACACAACACATTTGTGTCTG & 21 & 54	& 36 & 44\\ \hline
ACACAACACATTTGTGTCTGG & 32 & 61	& 44 & 31\\ \hline
CCTTGCACGTCTTGACAAAGT & 33 & 58	& 38 & 56\\ \hline
GCACGTCTTGACAAAGTTGAG & 32 & 22	& 45 & 32\\ \hline
Average Number & 35.166 & 52.000 & 51.000 & 45.75 \\
\hline
\end{tabular}
\caption{Numbers of reverse primer candidates generated under four feature extraction methods for each Alpha-variant forward primer.}
\label{tab:11}

{\raggedright \textbf{Note:} The number of reverse primer candidates varies with the feature extraction strategy. For each validated Alpha forward primer (left column), we report the number of reverse primer candidates produced by Pooling, Top-$k$, Mix, and Reconstruction. \par}

\end{table}

\begin{table}[H]
\centering
\small
\begin{tabular}{c|c|c|c|c}
Forward Primer (5$'$--3$'$) &	Delta &	Delta &	Delta &	Delta \\
 & Pooling Method & Top method & Mix Method & Recon Method \\ \hline
GATCACCGGTGGAATTGCTAC	& 16	& 68	& 29 & 39\\ \hline
GAATTGCTACCGCAATGGCTT	& 15	& 55	& 25 & 28\\ \hline
ACTCAGACTAATTCTCGTCGG	& 29	& 62	& 35 & 32\\ \hline
CTACCGCAATGGCTTGTCTTG	& 18	& 49	& 23 & 43\\ \hline
TGGAATTGCTACCGCAATGGC	& 17	& 60	& 29 & 23\\ \hline
CTCAGACTAATTCTCGTCGGC	& 27	& 70	& 41 & 26\\ \hline
ATTGCTACCGCAATGGCTTGT	& 20	& 60	& 24 & 27\\ \hline
AGACTCAGACTAATTCTCGTC	& 27	& 60	& 42 & 32\\ \hline
TAGATTTTGTTCGCGCTACTG	& 18	& 53	& 30 & 33\\ \hline
CCGGTGGAATTGCTACCGCAA	& 16	& 70	& 23 & 48\\ \hline
ACCGCAATGGCTTGTCTTGTA	& 17	& 74	& 22 & 32\\ \hline
GGTGGAATTGCTACCGCAATG	& 16	& 63	& 28 & 45\\ \hline
CTCCTTTAGATTTTGTTCGCG	& 26	& 65	& 16 & 25\\ \hline
TCACCGGTGGAATTGCTACCG	& 7	    & 60	& 24 & 2\\ \hline
ACCGGTGGAATTGCTACCGCA	& 18	& 54	& 24 & 38\\ \hline
ATCACCGGTGGAATTGCTACC	& 20	& 59	& 25 & 43\\ \hline
TACCGCAATGGCTTGTCTTGT	& 22	& 67	& 18 & 27\\ \hline
GGAATTGCTACCGCAATGGCT	& 14	& 66	& 28 & 31\\ \hline
CCGCAATGGCTTGTCTTGTAG	& 21	& 61	& 31 & 30\\ \hline
GTGGAATTGCTACCGCAATGG	& 15	& 64	& 22 & 25\\ \hline
TTGCTACCGCAATGGCTTGTC	& 16	& 66	& 31 & 43\\ \hline
GCTACCGCAATGGCTTGTCTT	& 10	& 66	& 27 & 22\\ \hline
TCAGACTCAGACTAATTCTCG	& 27	& 67	& 47 & 65\\ \hline
AATTGCTACCGCAATGGCTTG	& 15	& 60	& 29 & 22\\ \hline
CAGACTCAGACTAATTCTCGT	& 18	& 77	& 39 & 34\\ \hline
CGGTGGAATTGCTACCGCAAT	& 14	& 57	& 24 & 25\\ \hline
TGCTACCGCAATGGCTTGTCT	& 19	& 72	& 16 & 36\\ \hline
GACTCAGACTAATTCTCGTCG	& 26	& 74	& 44 & 26\\ \hline
Average Number & 18.714 & 63.536 & 28.429 & 32.214\\
\hline
\end{tabular}
\caption{Numbers of reverse primer candidates generated under four feature extraction methods for each Delta-variant forward primer.}
\label{tab:12}

{\raggedright \textbf{Note:} The number of reverse primer candidates varies with the feature extraction strategy. For each validated Delta forward primer (left column), we report the number of reverse primer candidates produced by Pooling, Top-$k$, Mix, and Reconstruction. \par}

\end{table}

\begin{table}[H]
\centering
\small
\begin{tabular}{c|ccccc}
\multirow{2}*{Reverse primer (5$'$--3$'$)} & SARS-CoV-2 & SARS-CoV-2 & SARS-CoV-2 & SARS-CoV-2 & SARS-CoV-2\\
~ &(Alpha) &(Beta) &(Gamma) & (Delta) &(Omicron)\\
\hline
\multirow{2}*{Dataset} & GISAID & GISAID  & GISAID  & GISAID  & GISAID \\
~ & and NCBI & and NCBI & and NCBI & and NCBI & and NCBI\\
Host & \textit{Homo sapiens} & \textit{Homo sapiens} & \textit{Homo sapiens} & \textit{Homo sapiens} & \textit{Homo sapiens} \\
Sequence Number & 5,000 & 5,000 & 5,000 & 5,000 & 5,000\\ \hline
Alpha Variant & & & & & \\ \hline
CATCAATGTCTCTGCCAAATTG & 0.9936 & 0.0004 & 0.0008 & 0.0002 & 0.0 \\
ACCAGACACAAATGTGTTGTGT & 0.9928 & 0.0008 & 0.0018 & 0.0004 & 0.0 \\
CAGACACAAATGTGTTGTGTG & 0.992 & 0.0008 & 0.0018 & 0.0004 & 0.0 \\
ACAGCATCAGTAGTGTCATCA & 0.9918 & 0.0004 & 0.0008 & 0.0002 & 0.0 \\ \hline
Beta Variant & & & & & \\ \hline
ACAGGGTTAGCAAACCTCTT & 0.0 & 0.9638 & 0.0 & 0.0006 & 0.0 \\
CTACTGCTGCCTGGAGTTG & 0.0016 & 0.9594 & 0.0004 & 0.0006 & 0.0008 \\
GTTCGTTTAGACCAGAAGATCAAG & 0.0002 & 0.9546 & 0.0 & 0.0004 & 0.0 \\
GGTTATGATTTTGGAAGCGCTA & 0.0006 & 0.953 & 0.0178 & 0.0012 & 0.0004 \\ \hline
Gamma Variant & & & & & \\ \hline
AATTTGGTCATCTCGACTG & 0.0 & 0.0 & 0.9918 & 0.0002 & 0.0 \\
TGGTCATCTCGACTGCTATTGGTGT & 0.0 & 0.0 & 0.9914 & 0.0002 & 0.0 \\
GCCAATTTGGTCATCTCGAC & 0.0 & 0.0 & 0.9912 & 0.0002 & 0.0 \\
TGAACCGTCGATTGTGTGAA & 0.0& 0.0006 & 0.985 & 0.0002 & 0.0 \\ \hline
Delta Variant & & & & & \\ \hline
CATTGCGGTAGCAATTCCA & 0.0006 & 0.0002 & 0.0 & 0.9952 & 0.165 \\
AGCGCGAACAAAATCTAAAGGA & 0.001 & 0.003 & 0.0006 & 0.9934 & 0.1634 \\
GTAGCGCGAACAAAATCTAAAGGAG & 0.0008 & 0.003 & 0.0006 & 0.9932 & 0.163 \\
GACGAGAATTAGTCTGAGTCTGAT & 0.0032 & 0.0004 & 0.0004 & 0.9896 & 0.1624 \\ \hline
Omicron Variant & & & & & \\ \hline
ATTGTGCCAACCACCATAGAA & 0.0038 & 0.0008 & 0.002 & 0.0036 & 0.834 \\
GGTGTGACTGTTATTGCCTGACCA & 0.0 & 0.0 & 0.0 & 0.0 & 0.8292 \\
TAACGTACCTATCTCTTCCGAA & 0.0032 & 0.0 & 0.001 & 0.0 & 0.8286 \\
CACCTGTGCCTTTTAAACCATTG & 0.0 & 0.0 & 0.0 & 0.0002 & 0.828 \\
\hline
\end{tabular}
\caption{Appearance frequencies of selected reverse primers across SARS-CoV-2 variants, evaluated on 5{,}000 \textit{Homo sapiens} host sequences per variant (combined GISAID and NCBI datasets).}
\label{tab:5.5}

{\raggedright \textbf{Note:} Values report the fraction of sequences in which each reverse primer occurs within the corresponding variant-specific dataset. Reverse primers are designed to be variant-discriminative; consequently, they show high occurrence in the target variant and low occurrence in non-target variants. \par}

\end{table}

\section{Flowcharts of the methodology}

\begin{figure}[H]
\centering
\includegraphics[width=0.5\textwidth]{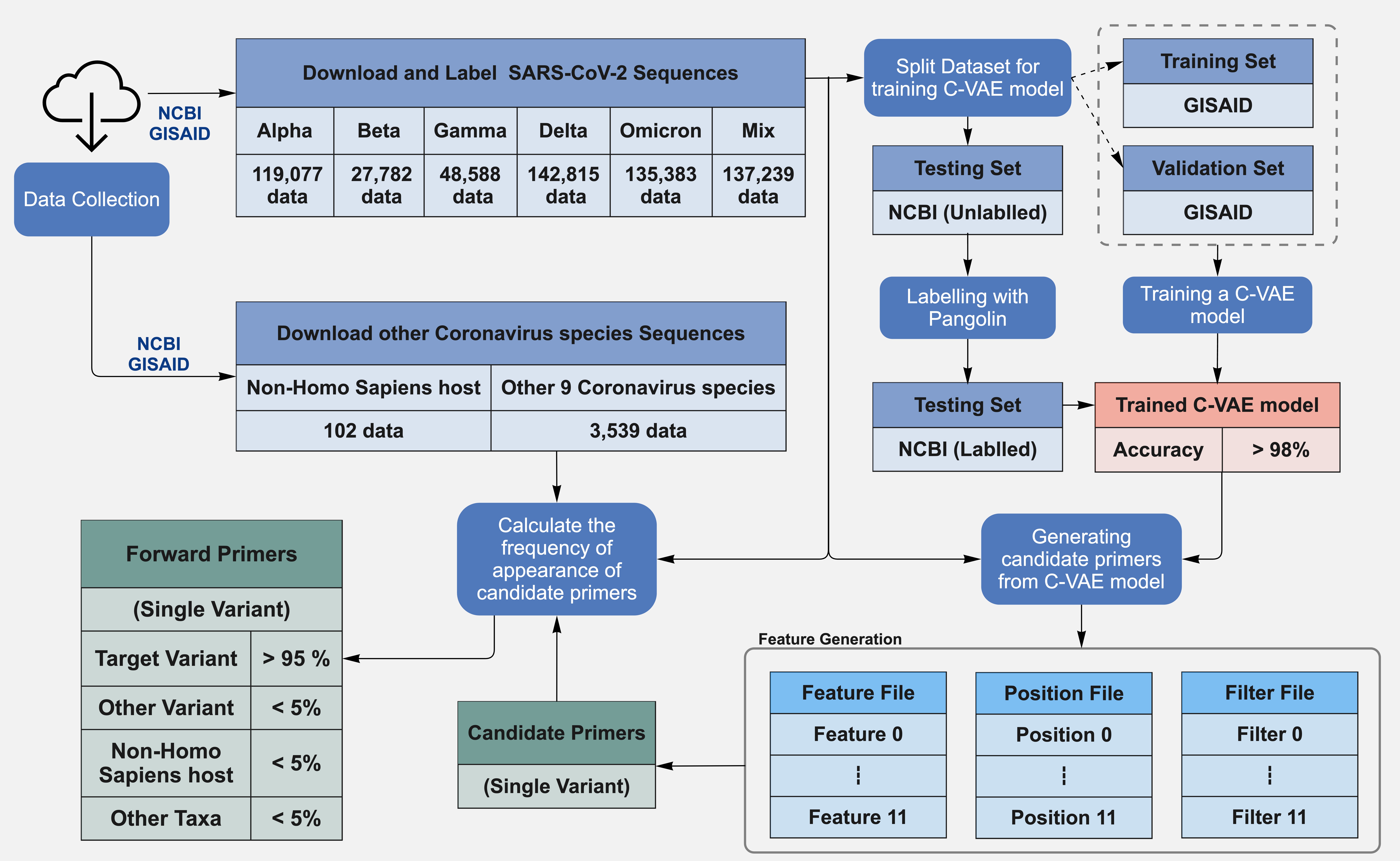}
\caption{Flowchart of the forward primer design workflow.}
\label{Flowchart_Forward}
\end{figure}

\begin{figure}[H]
\centering
\includegraphics[width=0.5\textwidth]{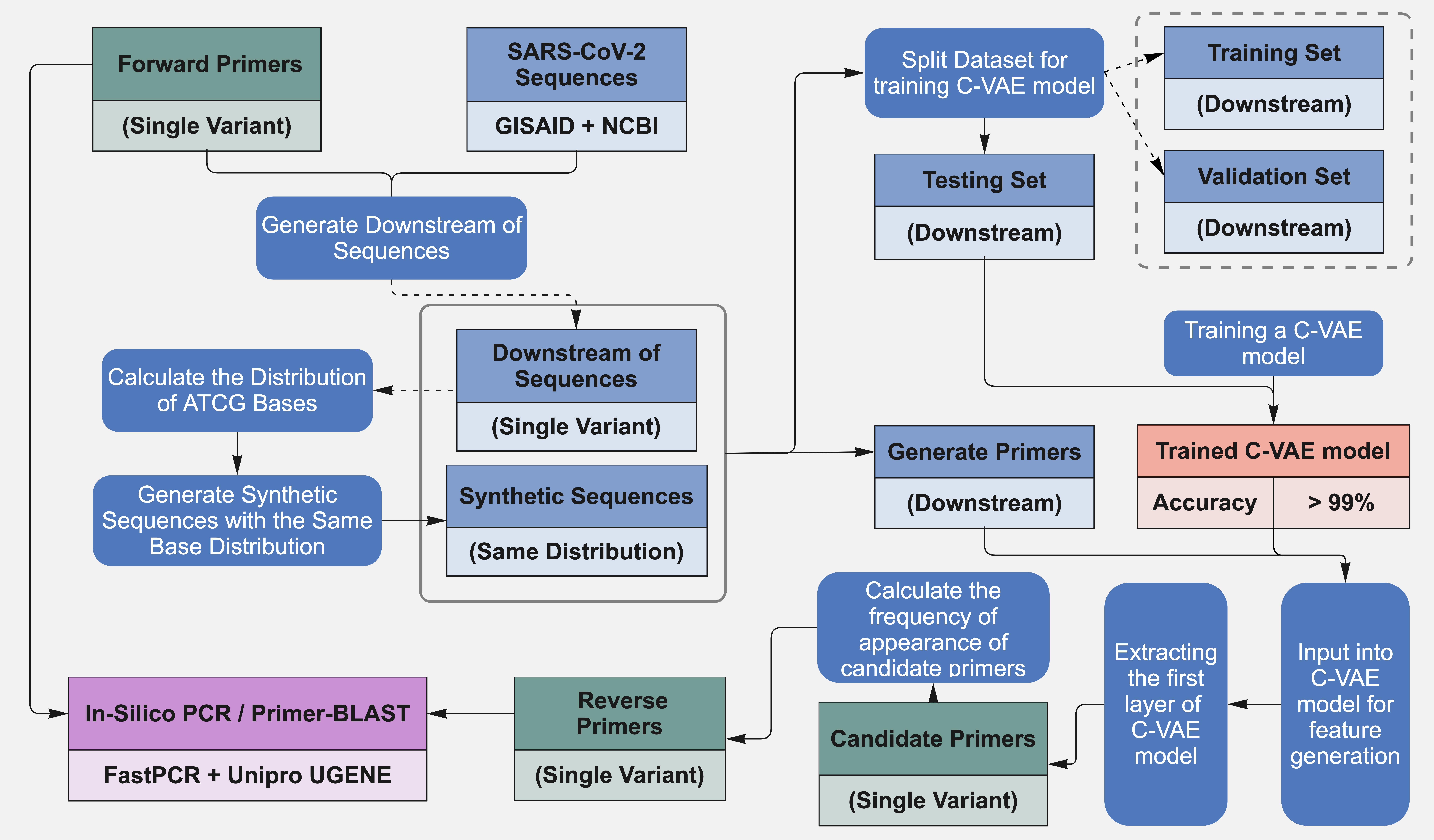}
\caption{Flowchart of the reverse primer design workflow.}
\label{Flowchart_Reverse}
\end{figure}

This section summarizes the complete workflows for forward and reverse primer design. Figure~\ref{Flowchart_Forward} presents the forward primer design pipeline, from data collection and preprocessing through feature extraction and screening to the final set of candidate forward primers. Figure~\ref{Flowchart_Reverse} illustrates the subsequent reverse primer design workflow, which is performed after forward primer selection. For an overview of the full end-to-end approach, see Figure~\ref{fig:overall_pipelines}.

\section{BLAST and \textit{in silico} PCR for primer validation} \label{Appendix_4}

\begin{figure}[H]
\centering
\includegraphics[width=0.4\textwidth]{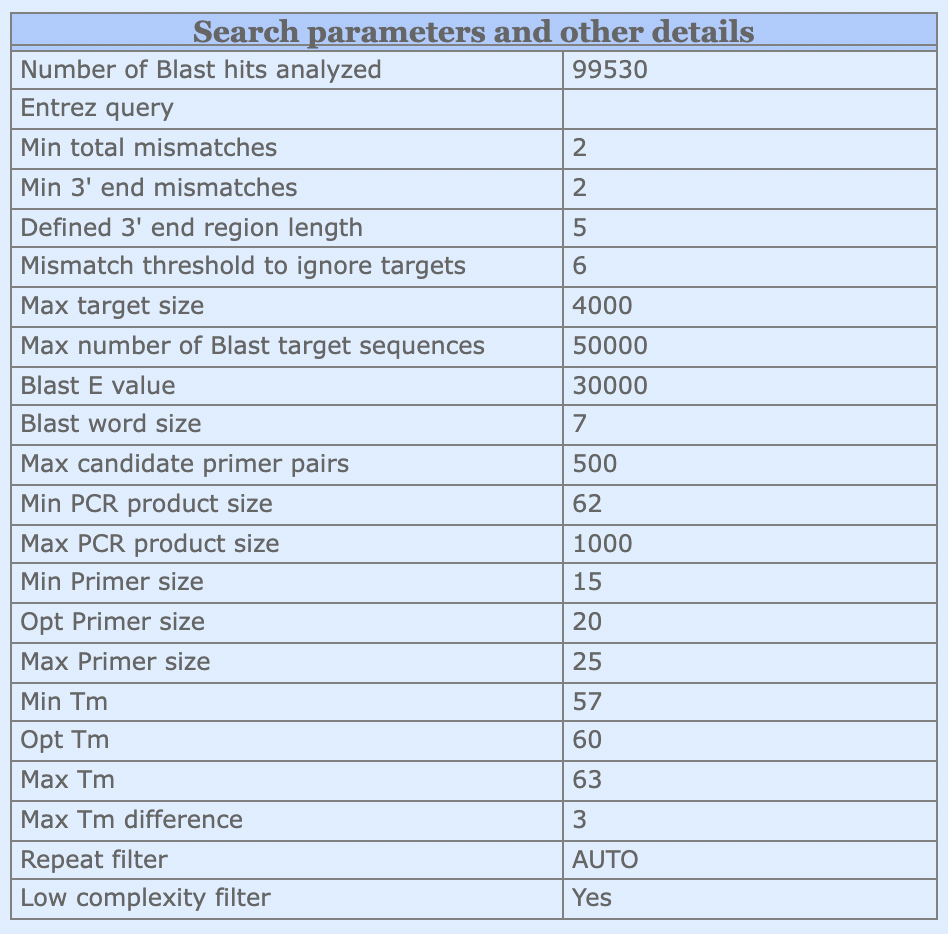}
\caption{Primer-BLAST search settings used for primer-pair validation.}
\label{fig:BALST}

{\raggedright \textbf{Note:} Primer-BLAST was run on the NCBI website with the search restricted to SARS-CoV-2 sequences. Key parameters were specified to ensure a consistent and valid specificity assessment. \par}
\end{figure}

\begin{figure}[H]
\centering
\includegraphics[width=0.5\textwidth]{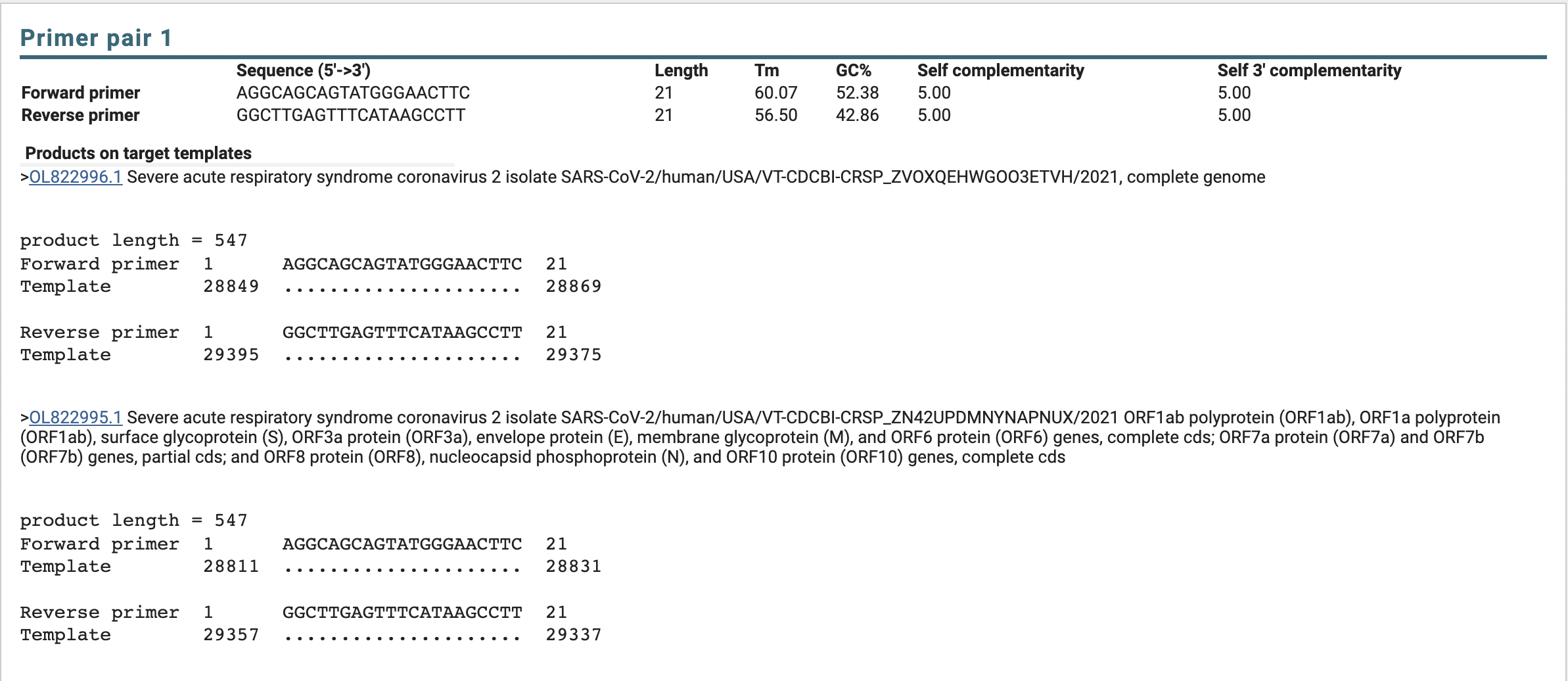}
\caption{Example Primer-BLAST output for primer-pair specificity assessment.}
\label{fig:BLAST_result}

{\raggedright \textbf{Note:} The output lists matching SARS-CoV-2 accessions identified by Primer-BLAST and reports the predicted amplicon length for each match under the specified settings. \par}
\end{figure}

This section provides representative outputs from Primer-BLAST, which we used to verify primer-pair specificity within SARS-CoV-2. Figures~\ref{fig:BALST} and \ref{fig:BLAST_result} show the Primer-BLAST search settings and an example of the resulting alignments, respectively.

\begin{figure}[H]
\centering
\includegraphics[width=0.5\textwidth]{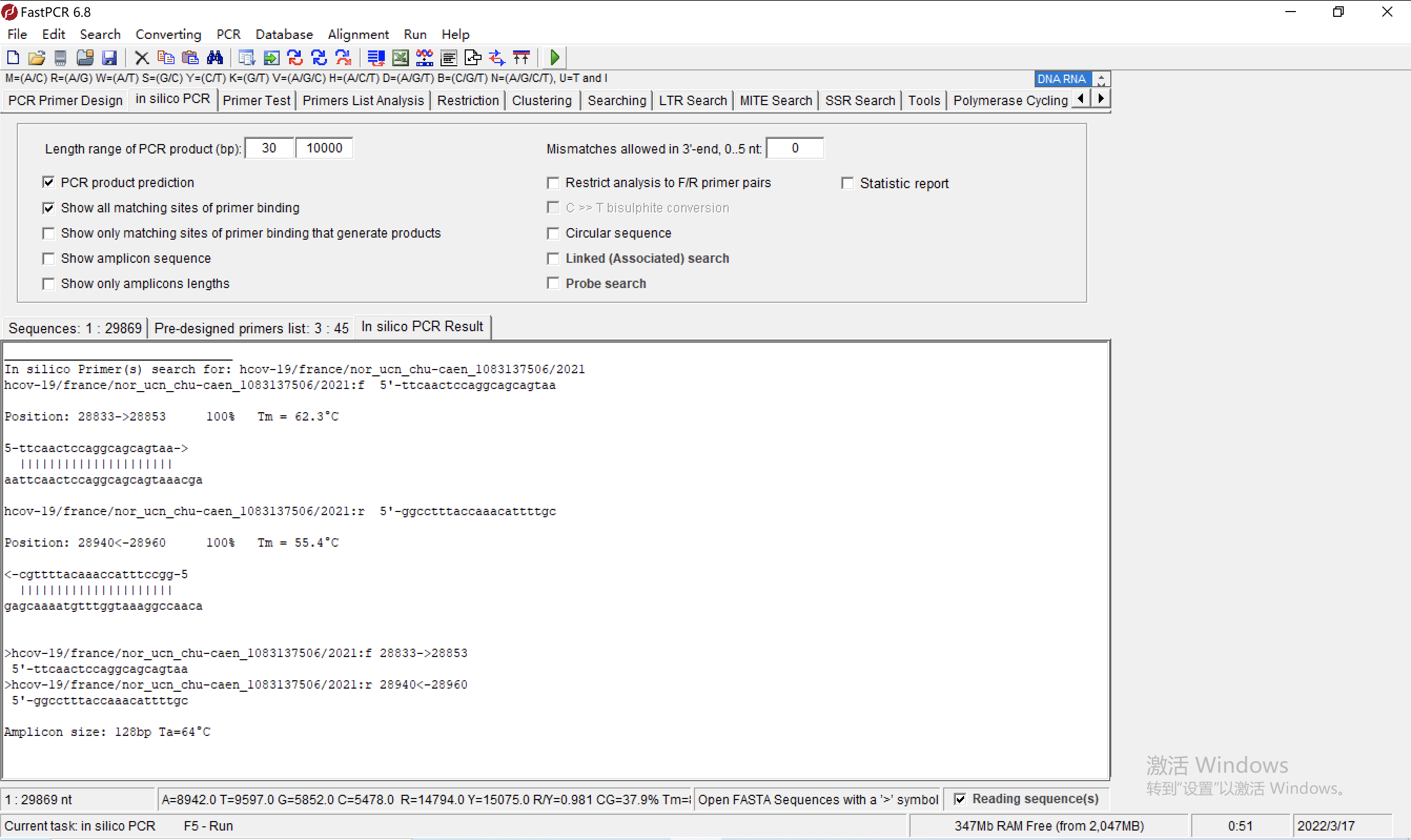}
\caption{Representative \textit{in silico} PCR results for selected primer pairs evaluated using FastPCR.}
\label{fig:FastPCR}
\end{figure}

\begin{figure}[H]
\centering
\includegraphics[width=0.5\textwidth]{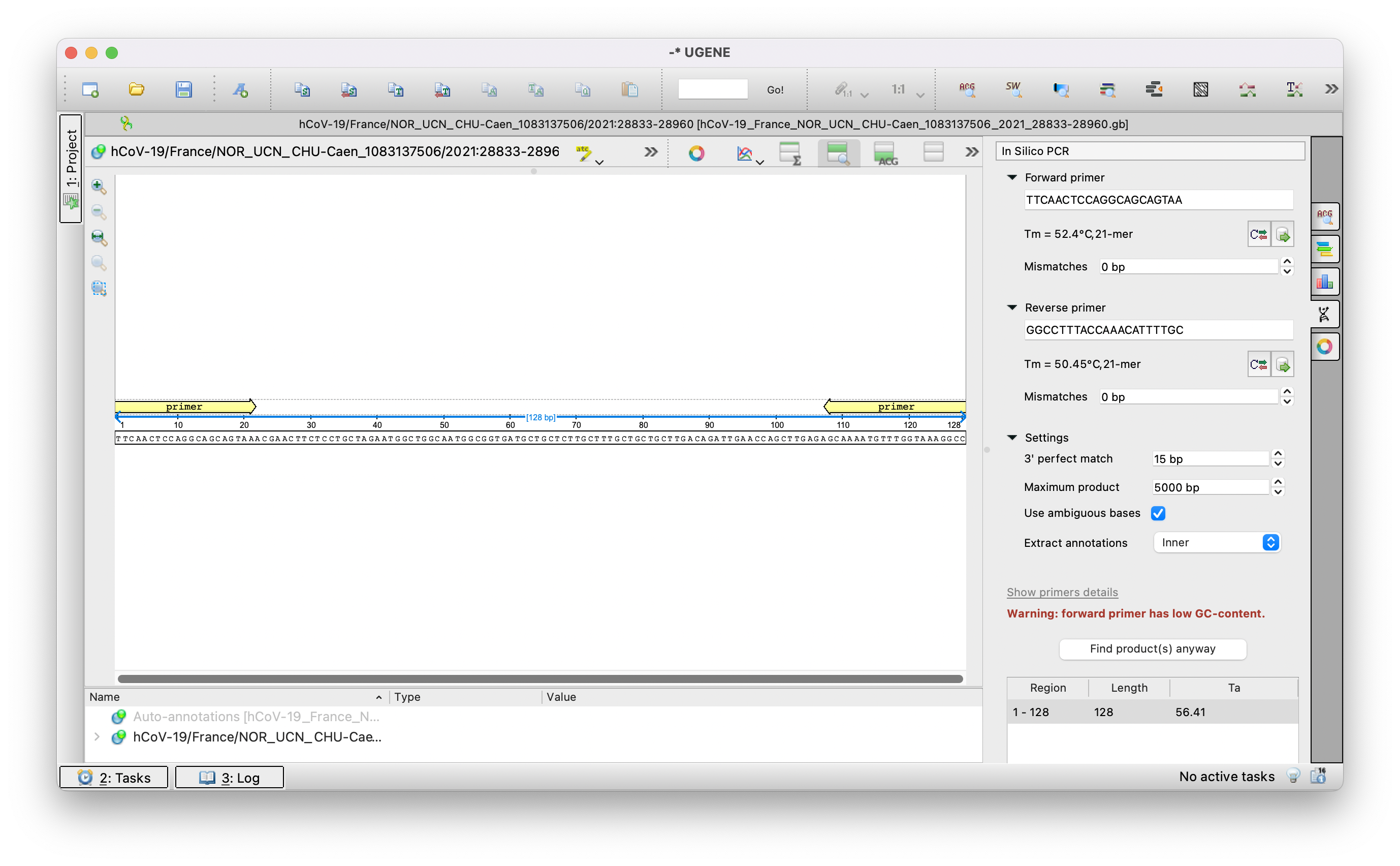}
\caption{Representative \textit{in silico} PCR results for selected primer pairs evaluated using Unipro UGENE.}
\label{fig:Unipro_UGENE}
\end{figure}

In this section, Figures~\ref{fig:FastPCR} and \ref{fig:Unipro_UGENE} show representative \textit{in silico} PCR results obtained using FastPCR and Unipro UGENE, respectively.

\section{Comparison with Primer3Plus} \label{Appendix_3}

This appendix compares Primer C-VAE with Primer3Plus. Table~\ref{tab: Primer3_vs_C-VAE} summarizes primer pairs generated by our deep-learning pipeline (including both forward and reverse primers) and contrasts them with primer pairs generated using Primer3Plus under the corresponding settings.

\begin{table}[H]
\centering
\small
\begin{tabular}{ccccc}
\multicolumn{1}{c}{} & \multicolumn{2}{c}{\textbf{Primer C-VAE}} & \multicolumn{2}{c}{\textbf{Primer3Plus}} \\ \hline
\multicolumn{1}{c}{} & \multicolumn{2}{c}{(No practical upper limit on input length)} & \multicolumn{2}{c}{(Up to 10{,}000 bp per input)} \\ \hline
\textbf{Forward primers} & \multicolumn{4}{c}{} \\ \hline
\multicolumn{5}{c}{} \\
\multicolumn{1}{c|}{\textit{Total (initial)}} & 3{,}626 primers  & \multicolumn{1}{c|}{} & 30 primers & \\
\multicolumn{1}{c|}{\textit{GC-content filtering}} & 2{,}725 primers & \multicolumn{1}{c|}{$\downarrow$24.85\%} & 30 primers & \multicolumn{1}{c}{$\downarrow$0\%} \\
\multicolumn{1}{c|}{\textit{Appearance-frequency filtering}} & 29 primers & \multicolumn{1}{c|}{$\downarrow$98.94\%}  & 0 primers & \multicolumn{1}{c}{$\downarrow$100\%} \\
\multicolumn{5}{c}{} \\ \hline

\textbf{Reverse primers} & \multicolumn{4}{c}{} \\ \hline
\multicolumn{5}{c}{} \\
\multicolumn{1}{c|}{\textit{Total (initial)}} & 53{,}777 primers  & \multicolumn{1}{c|}{} & 30 primers & \\
\multicolumn{1}{c|}{\textit{GC-content filtering}} & 34{,}955 primers & \multicolumn{1}{c|}{$\downarrow$35\%} & 30 primers & \multicolumn{1}{c}{$\downarrow$0\%} \\
\multicolumn{1}{c|}{\textit{Appearance-frequency filtering}} & 357 primers & \multicolumn{1}{c|}{$\downarrow$98.98\%}  & 0 primers & \multicolumn{1}{c}{$\downarrow$100\%} \\
\multicolumn{5}{c}{} \\ \hline

\textbf{Primer pairs} & \multicolumn{4}{c}{} \\ \hline
\multicolumn{5}{c}{} \\
\multicolumn{1}{c|}{\textit{Primer design rules}} & \multicolumn{2}{c|}{143 primer pairs} & \multicolumn{2}{c}{0 primer pairs} \\
\multicolumn{5}{c}{} \\ \hline
\end{tabular}
\caption{Comparison between Primer C-VAE and Primer3Plus for primer generation and filtering outcomes.}
\label{tab: Primer3_vs_C-VAE}

{\raggedright \textbf{Note:} Primer3Plus limits the input sequence length to 10{,}000~bp and returns up to 10 primer pairs per input (10 forward and 10 reverse). For SARS-CoV-2 (\textasciitilde30{,}000~bp), the genome must therefore be split into three segments and processed separately, yielding 30 candidate primers of each type before downstream filtering. \par}
\end{table}

\end{appendices}

\end{document}